\documentclass[man]{apa7}
\usepackage{enumitem}
\usepackage{newfloat}
\DeclareFloatingEnvironment[name={Supplementary Figure}]{suppfigure}

\renewcommand\appendixname{Supplimentary Information}
\usepackage[utf8]{inputenc}
\usepackage{authblk}
\usepackage{setspace}
\usepackage{graphicx}
\usepackage{subcaption}
\usepackage{amsmath}
\usepackage{lineno}
\usepackage{url}
\DeclareGraphicsExtensions{.png,.pdf}
\usepackage{algorithm2e}
\usepackage{longtable}
\usepackage{xcolor}
\usepackage{appendix}
\usepackage{ragged2e}
\usepackage{hyperref}

\usepackage[style=apa,sortcites=true,sorting=nyt,backend=biber]{biblatex}
\title{Analyzing time series of unequal durations using Multidimensional Recurrence Quantification Analysis (MdRQA): validation and implementation using Python}
\shorttitle{Sliding window MdRQA validation}
\affiliation{Swarag Thaikkandi\textsuperscript{1, 2$\dag$}, K. M. Sharika\textsuperscript{2*} \\ \textsuperscript{1}Department of Biology, Indian Institute of Science Education and Research, Pune, Maharashtra, India \\ \textsuperscript{2}Department of Cognitive Science, Indian Institute of Technology Kanpur, Kanpur, Uttar Pradesh, India}
\authornote{
   $\dag$ Current affiliation \\
   * Address correspondence to }

\abstract{In recent years, recurrent quantification analysis (RQA) and its multi-dimensional version (MdRQA) have emerged as a popular tool for assessing interpersonal behavioral or physiological synchrony in groups of two or more individuals. While experimental data in such studies are typically collected for a fixed, pre-determined duration, naturally occurring phenomena may often reach a state of transition after an unpredictable or varying duration of time. The resulting recurrence plots (RPs) across samples cannot be compared directly via linear scaling because the sensitivity of RQA variables to local dynamics would vary. We propose to address this by using the sliding window technique on individual RPs and using the summary statistics of the different RQA variable distributions computed across the sliding windows to differentiate the dynamics of the original time series of unequal durations. We first tested our approach in two simulated models: 1) the Rossler attractor and  2) the Kuramoto model. We compared the ability of different summary statistics of RQA variable distributions to accurately predict the dynamic states of the system across varying levels of noise, unequal lengths of time series, and, in the case of the Kuramoto model, different numbers of oscillators across samples. We found that while the mean, compared to other measures of central tendency, was a more accurate predictor of the underlying dynamic state of the system at high noise conditions, the mode was the most robust to the degree of noise in the signals. To demonstrate the efficacy of the sliding window MdRQA approach in real world experimental settings, we further tested and validated it on open access data from a recent study comparing spontaneously generated interpersonal movement synchrony in a dyad, across visual access and proximity conditions. To our knowledge, this is the first systematic attempt to validate the use of MdRQA in computing and comparing synchrony between systems of non-uniform composition and unequal time series data, paving the way for future work that examines interpersonal synchrony in more naturalistic, ecologically valid contexts. }

\begin{document}
\maketitle
\section{Introduction}
\justifying \par Recent years have witnessed a steady rise in the number of studies examining behavioral and/ or physiological synchrony between individuals in relation to their interpersonal outcomes including cohesion, cooperation and trust (\citeauthor{guastello2006electrodermal}, \citeyear{guastello2006electrodermal}; \citeauthor{richardson2007art}, \citeyear{richardson2007art}; \citeauthor{shockley2003mutual}, \citeyear{shockley2003mutual}; \citeauthor{schippers2010mapping}, \citeyear{schippers2010mapping}; \citeauthor{konvalinka2011synchronized}, \citeyear{konvalinka2011synchronized}; \citeauthor{yun2012interpersonal}, \citeyear{yun2012interpersonal}; \citeauthor{hasson2012brain}, \citeyear{hasson2012brain}; \citeauthor{spiegelhalder2014interindividual}, \citeyear{spiegelhalder2014interindividual}; \citeauthor{golland2015mere}, \citeyear{golland2015mere}; \citeauthor{mitkidis2015building}, \citeyear{mitkidis2015building}; \citeauthor{wallot2016multidimensional}, \citeyear{wallot2016multidimensional}; \citeauthor{bevilacqua2019brain}, \citeyear{bevilacqua2019brain}; \citeauthor{bevilacqua2019brain}, \citeyear{bevilacqua2019brain}; \citeauthor{pan2020instructor}, \citeyear{pan2020instructor}; \citeauthor{thorson2021oxytocin}, \citeyear{thorson2021oxytocin}; \citeauthor{gordon2021group}, \citeyear{gordon2021group}; \citeauthor{tomashin2022interpersonal},\citeyear{tomashin2022interpersonal}). Recurrent quantification analysis (RQA) has emerged as a popular tool for assessing interpersonal synchrony in such studies primarily since it makes very few assumptions about the constituent components of the system under study and because of its general robustness to dynamic variability in the signal (also known as non-stationarity) (\citeauthor{marwan2007recurrence}, \citeyear{marwan2007recurrence}). RQA is based on the principle that in a complex system governed by many inter-dependent, interacting components, any measured variable or dimension (and  its time delayed copies) could be used to retrieve the dynamical behavior  of the system or in other words, reconstruct what is known as its \emph{phase space} or \emph{state space} (\citeauthor{takens1981dynamical}, \citeyear{takens1981dynamical}). Phase space is a graphical representation of all possible states of the system over time and is composed of all the coordinates (or dimensions) required to define a given state of the system (\citeauthor{huffaker2017nonlinear}, \citeyear{huffaker2017nonlinear}). Quantifying \emph{recurrence} or how often a trajectory revisits a point in the phase space (typically represented as black dots against a white background in a recurrence plot, Fig.\ref{fig:4}) would indicate whether and how different components of the multi-variate system interact and converge on the same state (or in other words, are \emph{in sync}) across time (\citeauthor{webber1994dynamical}, \citeyear{webber1994dynamical}; \citeauthor{marwan2002recurrence}, \citeyear{marwan2002recurrence}). MdRQA or multidimensional RQA is an extension of RQA that allows the time series data of more than two components (or dimensions) of the system to be considered simultaneously for computing recurrence at the group level (\citeauthor{wallot2016multidimensional}, \citeyear{wallot2016multidimensional}; \citeauthor{gordon2021group}, \citeyear{gordon2021group}; \citeauthor{tomashin2022interpersonal}, \citeyear{tomashin2022interpersonal}). 
\par Recent years have witnessed a steady rise in the number of studies examining behavioral and/or physiological synchrony between individuals in relation to their interpersonal outcomes including cohesion, cooperation and trust (\citeauthor{guastello2006electrodermal}, \citeyear{guastello2006electrodermal}; \citeauthor{richardson2007art}, \citeyear{richardson2007art}; \citeauthor{shockley2003mutual}, \citeyear{shockley2003mutual}; \citeauthor{schippers2010mapping}, \citeyear{schippers2010mapping}; \citeauthor{konvalinka2011synchronized}, \citeyear{konvalinka2011synchronized}; \citeauthor{yun2012interpersonal}, \citeyear{yun2012interpersonal}; \citeauthor{hasson2012brain}, \citeyear{hasson2012brain}; \citeauthor{spiegelhalder2014interindividual}, \citeyear{spiegelhalder2014interindividual}; \citeauthor{golland2015mere}, \citeyear{golland2015mere}; \citeauthor{mitkidis2015building}, \citeyear{mitkidis2015building}; \citeauthor{wallot2016multidimensional}, \citeyear{wallot2016multidimensional}; \citeauthor{bevilacqua2019brain}, \citeyear{bevilacqua2019brain}; \citeauthor{dikker2017brain}, \citeyear{dikker2017brain}; \citeauthor{pan2020instructor}, \citeyear{pan2020instructor}; \citeauthor{thorson2021oxytocin}, \citeyear{thorson2021oxytocin}; \citeauthor{gordon2021group}, \citeyear{gordon2021group}; \citeauthor{tomashin2022interpersonal},\citeyear{tomashin2022interpersonal}). Recurrent quantification analysis (RQA) has emerged as a popular tool for assessing interpersonal synchrony in such studies primarily since it makes very few assumptions about the constituent components of the system under study and because of its general robustness to dynamic variability in the signal (also known as non-stationarity) (\citeauthor{marwan2007recurrence}, \citeyear{marwan2007recurrence}). RQA is based on the principle that in a complex system governed by many inter-dependent, interacting components, any measured variable or dimension (and  its time delayed copies) could be used to retrieve the dynamical behavior  of the system or in other words, reconstruct what is known as its \emph{phase space} or \emph{state space} (\citeauthor{takens1981dynamical}, \citeyear{takens1981dynamical}). Phase space is a graphical representation of all possible states of the system over time and is composed of all the coordinates (or dimensions) required to define a given state of the system (\citeauthor{huffaker2017nonlinear}, \citeyear{huffaker2017nonlinear}). Quantifying \emph{recurrence} or how often a trajectory revisits a point in the phase space (typically represented as black dots against a white background in a recurrence plot, Fig.\ref{fig:4}) would indicate whether and how different components of the multi-variate system interact and converge on the same state (or in other words, are \emph{in sync}) across time (\citeauthor{webber1994dynamical}, \citeyear{webber1994dynamical}; \citeauthor{marwan2002recurrence}, \citeyear{marwan2002recurrence}; \citeauthor{mitkidis2015building}, \citeyear{mitkidis2015building}). MdRQA or multidimensional RQA is an extension of RQA that allows the time series data of more than two components (or dimensions) of the system to be considered simultaneously for computing recurrence at the group level (\citeauthor{wallot2016multidimensional}, \citeyear{wallot2016multidimensional}; \citeauthor{gordon2021group}, \citeyear{gordon2021group}; \citeauthor{tomashin2022interpersonal}, \citeyear{tomashin2022interpersonal}). 
\justifying \par While experimental data assessing synchrony between multiple components of a system are typically collected for a fixed, pre-determined duration, naturally occurring phenomena may often reach a state of transition after an unpredictable or varying duration of time at which point data collection may be discontinued. This results in time series data of unequal lengths across different samples of the system. For example, it is possible that when one is recording behavioral or physiological variables simultaneously from two or more individuals in a group, different groups being sampled for the study successfully complete the task or arrive at a meaningful interpersonal outcome at varied lengths of time. It may also be the case that, for some logistic reason, data could not be collected for a uniform duration across all groups. In such scenarios, the resulting recurrence plots (RPs) of different sizes cannot be compared directly via linear scaling because RQA variables are statistical measures of the vertical or horizontal line distributions in an RP  and consequently,  represent the local dynamical changes to varying levels of sensitivity based on the overall length of the individual RP. In other words, while RQA variables from shorter RPs may be able to capture local dynamics at relatively high resolution, the same may get averaged out in larger RPs. This makes the comparison of synchrony measures across groups difficult, particularly given recent reports demonstrating the impact of total pooled duration of data (in minutes) on detection of correlation based synchrony across participants (\citeauthor{stuldreher2023robustness}, \citeyear{stuldreher2023robustness}). We proposed to address this by applying the sliding window technique on RPs, previously shown to be demonstrably sensitive for tracking local dynamical changes in physiology (\citeauthor{webber2005recurrence}, \citeyear{webber2005recurrence}; \citeauthor{webber1995influence}, \citeyear{webber1995influence}). The sliding window technique involves partitioning a single large RP into smaller, same sized, overlapping windows along the RP diagonal and then computing the RQA variables for each of these sliding windows. We hypothesized that the summary statistics of the different RQA variable distributions computed across the sliding windows of a given RP could be used to functionally differentiate the dynamics of a set of such RPs using a nested cross-validation approach (Fig. \ref{fig:1}).
\justifying \par We proceeded to examine this hypothesis by first using two well-known  non-linear dynamical systems as our testing grounds: 1) Rossler attractor and  2) the Kuramoto system of coupled oscillators. We tested the ability of our approach to classify the dynamics of each system with high cross validation accuracy – while the duration of the time series, and in the case of the Kuramoto model, the number of oscillators (or observed dimensions) in the system varied across samples. We also tested the robustness of different summary statistics of MdRQA variable distributions (across sliding windows of a RP) in accurately predicting the dynamic states of the system across varying levels of noise in the signal. To further validate the efficacy of our approach in complex, real-world scenarios where absolute levels of noise are unknown, we tested it on openly accessible data from \citeauthor{koul2023interpersonal}(\citeyear{koul2023interpersonal}), examining interpersonal movement synchrony between individuals in a dyad (across varying experimental conditions of visual access and proximity) and asked if our approach could replicate the trends reported in the original study.  

\subsection{Multi-dimensional Recurrence Quantification Analysis or MdRQA}
RQA was developed to quantify the dynamics of a time series (\citeauthor{webber1994dynamical}, \citeyear{webber1994dynamical}; \citeauthor{marwan2002recurrence}, \citeyear{marwan2002recurrence}) via phase space reconstruction using time delayed embedding (\citeauthor{takens1981dynamical}, \citeyear{takens1981dynamical}). As described above, phase-space is a mathematical construct that serves as a graphical representation to comprehensively depict the complete range of states that a system being analyzed can occupy. If the complete depiction of the system requires it to have D independent measures, then the phase-space is said to have D dimensions. Time delayed embedding is a method that aims to retrieve the dynamics of a multi-component, coupled system by using the time series data of one of the interacting components which is assumed to encode the information about the dynamics of others in itself (\citeauthor{hasson2012brain}, \citeyear{hasson2012brain}). According to Taken’s theorem, if a system consists of multiple, independent, interacting dimensions, but, if one has access to only one of them over time (say, x), then the dynamics of the whole system can be approximately reconstructed (a process known as embedding) by utilizing (D number of) time delayed versions of the observable x as (D-dimensional) coordinates of the phase space (\citeauthor{huffaker2017nonlinear},\citeyear{huffaker2017nonlinear}). For example, let the time series be:

\begin{equation}
    \boldsymbol{x} = ( x_{1}, x_{2}, x_{3}, ..., x_{n})
\end{equation}
which is sampled at regular intervals of time. From this, D-dimensional vectors, $\boldsymbol{V_{1}}$, $\boldsymbol{V_{2}}$, and so on, can be constituted by estimating D-1  versions of $\boldsymbol{x}$ with time delay, $\tau$, as below:
\begin{equation}
    \boldsymbol{V_{1}}= (x_{1}, x_{1+\tau}, x_{1+2\tau}, ..., x_{1+(D-1)\tau})
\end{equation}
\begin{equation}
    \boldsymbol{V} = \begin{bmatrix}
                        \boldsymbol{V_{1}} \\
                        \boldsymbol{V_{2}} \\
                        \vdots \\
                        \boldsymbol{V_{n-(D-1)\tau}}
                      \end{bmatrix}  =
                      \begin{bmatrix}
                          x_{1}, x_{1+\tau}, x_{1+2\tau}, ..., x_{1+(D-1)\tau} \\
                          x_{2}, x_{2+\tau}, x_{2+2\tau}, ..., x_{2+(D-1)\tau} \\
                          x_{3}, x_{3+\tau}, x_{3+2\tau}, ..., x_{3+(D-1)\tau} \\
                          \vdots \\
                          x_{n-(D-1)\tau}, x_{n-(D-2)\tau}, x_{n-(D-3)\tau}, ..., x_{n}
                      \end{bmatrix}
\end{equation}

In the above expression, while each row represents a point in the D-dimensional phase space, each column represents the coordinates in the evolution of the trajectory in one of the D dimensions.
\justifying \par A recurrence plot or RP generated by graphically representing how often a trajectory revisits a point in the phase space (\citeauthor{eckmann1995recurrence}, \citeyear{eckmann1995recurrence}) lends itself for statistical analysis that is quantified by the RQA variables. RP contains binary values denoting repetition of values in V over time. However, it is possible that the points may not always revisit exactly the same spot in the phase space in which case a predetermined threshold criterion, $\epsilon$, is used to classify whether a given point is close enough to be deemed recurrent or not. For example, if the euclidean distance between two points in the trajectory (at time points i and j, respectively) is less than $\epsilon$, a value of 1 (=recurrent) would be assigned to the ith row and jth column in the RP. This is accomplished by the heaviside step function, that would assign 1, if the value of input is greater than 0, and 0 otherwise. 
\begin{equation}
    RP_{ij}= \Theta(\epsilon - ||\boldsymbol{V_{i}} - \boldsymbol{V_{j}}||)
\end{equation}
MdRQA differs from RQA in that it allows multiple measures or observables to be used for phase space reconstruction (\citeauthor{wallot2016multidimensional}, \citeyear{wallot2016multidimensional}). Please see Supplementary section \ref{section:tde} for details on how expression (3) would transform in case of MdRQA.

\section{Methods}
\subsection{Parameter Estimation}
\justifying \par To generate the recurrence plots, time delay \& embedding dimension is computed for each dynamical system that has been observed or measured as part of the dataset. Time delay ($\tau$) is estimated by serially sampling time delays from 1 to 20 and computing the first minima (first local minima or global minima when the local minima did not exist) of the mutual information between the time series of a group and a time delayed version of it (\citeauthor{wallot2018calculation}, \citeyear{wallot2018calculation}; please see Supplementary section \ref{section:time-delay} for more details). This approach ensures that the time delayed signals are not too similar and permits the multidimensional topology of the trajectories in the phase space to unfold completely. While \citeauthor{wallot2016multidimensional}(\citeyear{wallot2016multidimensional}) computed this for each time series of the group (dimension) separately and averaged the value across all members of the group, to capture the cross information between different dimensions, we computed the multidimensional mutual information by estimating the multidimensional probability distributions for the group (please see section \ref{section:time-delay} Information for more details).

\begin{longtable}{|p{4cm}|p{5cm}|p{4cm}|}
\hline
\textbf{Measure} & \textbf{Formula} & \textbf{Description} \\
\hline
\center Recurrence Rate (RR) & \center$RR = \frac{1}{N^2}\sum_{i=1}^{N}\sum_{j=1}^{N} R(i,j)$ & Measures density of recurrence points in the recurrence plot, indicating how probable recurrence of states is in the system. \\
\hline
\center Determinism (DET) & \center$DET = \frac{\sum_{l=l_{\text{min}}}^{N}l\cdot p(l)}{\sum_{l=1}^{N}l\cdot p(l)}$ & Measures what fraction of the diagonal line lengths are above a minimum, given, line length. $p(l)$ is the probability of a line length $l$. Since diagonal lines are markers of consecutive periods of recurrence in the data, determinism corresponds to the predictability of the dynamical system. \\
\hline
\center Laminarity (LAM) & \center$LAM = \frac{\sum_{v=v_{\text{min}}}^{N}v\cdot p(v)}{\sum_{v=1}^{N}v\cdot p(v)}$ & Mathematically equivalent to determinism but defined for vertical (or horizontal) line lengths. Since vertical (or horizontal) lines are markers of states that do not change or change very slowly, laminarity quantifies the extent of the dynamical system being trapped in any given state for some time (\citeauthor{marwan2007recurrence}, \citeyear{marwan2007recurrence}). \\
\hline
\center Average diagonal line length (L) & \center$L = \frac{\sum_{l=l_{\text{min}}}^{N}l\cdot p(l)}{\sum_{l_{\text{min}}}^{N}p(l)}$ & Average value of diagonal line length distribution, quantifying how far in time the dynamical system is predictable. \\
\hline
\center Average vertical line length (Trapping time) (TT) & \center$TT = \frac{\sum_{v=v_{\text{min}}}^{N}v\cdot p(v)}{\sum_{v_{\text{min}}}^{N}p(v)}$ & Average value of the vertical line length distribution. \\
\hline
\center Maximum diagonal line length & & Maximum value from the diagonal line distribution \\
\hline
Maximum vertical line length & & Maximum value from the vertical line distribution \\
\hline
\center Diagonal and Vertical Entropy (ENTR) & \center$ENTR = -\sum_{l=l_{\text{min}}}^{N} p(l) \ln p(l)$ & Quantifies the degree of uncertainty in the possible states and hence, the complexity of the dynamical system, using the distribution of diagonal and vertical line lengths present in the plot, respectively. \\
\hline
\caption{Measures in Recurrence Analysis}
\label{table-1}
\end{longtable}

\justifying \par The number of embedding dimensions, $\boldsymbol{m}$, required to adequately reconstruct the phase space is estimated using the false nearest neighbor approach (\citeauthor{kennel1992determining}, \citeyear{kennel1992determining}; \citeauthor{hegger1999improved}, \citeyear{hegger1999improved}). False nearest neighbors (FNN) are points in the phase space that cease to be neighbors as the embedding dimension is increased and the ratio of the distances between them at the higher dimension ($\boldsymbol{m+1}$) and the current ($\boldsymbol{m}$) becomes larger than a threshold, $\boldsymbol{r}$. An increase in FNN, for a given choice of $\boldsymbol{r}$, is indicative of a phase space that needs to be reconstructed with more dimensions to unfold it completely. To arrive at an appropriate $\boldsymbol{r}$ for comparing FNN across embedding dimensions, we first plotted the FNN ratio for different values of $\boldsymbol{r}$ as embedding dimension was increased from 1 to 10. As embedding dimensions increased, we not only found FNN ratios to hit zero at smaller $\boldsymbol{r}$ values (Kantz \& Schreiber, 2003), but the values of $\boldsymbol{r}$, at which FNN ratios hit zero, also varied less at the higher embedding dimensions indicative of a potentially unfolded phase space. We set the tolerance criteria for the difference between the $\boldsymbol{r}$ at which FNN ratio at $\boldsymbol{m+1}$th dimension and $\boldsymbol{m}$ dimension hits zero to be 0.2 and plotted the corresponding $\boldsymbol{r}$ values as a function of embedding dimension (Supplementary Fig. \ref{fig:S2}) to select the m at which $\boldsymbol{r}$ was beginning to change the least (i.e. the knee point of the $\boldsymbol{r}$ vs. m plot). 
\justifying \par Threshold radius, $\epsilon$, which decides how close two points in the phase space should be to be considered recurrent, is chosen such that the recurrence rate (percentage of recurrent points/ black dots in the recurrence plot) is constant across different multi-component systems or RPs under study. Recurrence rate was kept constant at 10\% in our case allowing RQA variables derived from different samples of a dynamical system to be directly comparable by controlling for the degree of sparseness. 
\subsection{Comparing MdRQA variables across RPs of unequal lengths }
To arrive at the best approach for comparing MdRQA variables across RPs of unequal lengths, we used two well known dynamical systems as our testing grounds: the Rossler attractor and the Kuramoto system of coupled oscillators. While in the former case, we tested whether the summary statistics of MdRQA variable distributions (across sliding widows of each RP) could correctly classify between the chaotic vs. periodic states of a sample of Rossler attractors, in the latter case, we tested whether the summary statistics of MdRQA variable distributions (across sliding windows of each RP) could correctly classify if the coupling strength of the Kuramoto oscillators in each sample of the system was greater than the critical coupling strength, $\boldsymbol{K_{c}}$, or not. In both examples, the lengths of time series were allowed to vary across samples and data simulated across nine different levels of noise. In the Kuramoto system, the number of oscillators was randomly sampled too from a range of 3 to 6 oscillators.
\subsection{Selecting the appropriate sliding window size}
Since RQA variables are statistical measures (e.g., average, max, entropy etc.) obtained from histogram distributions of vertical or horizontal lines of different lengths in the RP, it is important that the time window being examined for generating these histogram distributions be large enough to estimate the local dynamics with low variance and high statistical confidence. To select the time window that can be used to compare RPs of unequal sizes, we used a bootstrapping method (\citeauthor{marwan2013recurrence}, \citeyear{marwan2013recurrence}) with sliding windows of different lengths to first estimate the lowest sliding window size likely to give MdRQA variable estimates with high statistical confidence in a given dataset. We tested sliding windows of different sizes (10 to 500, each window shifted but just one unit in time sampled for the time series) on all simulated RPs. For each sliding window, we first generated a histogram distribution of lines (vertical \& horizontal separately) of different lengths and then summed the counts corresponding to each bin of the histogram distributions from all the sliding windows to get a unified probability distribution for that window size (Supplementary Fig. \ref{fig:S3} \& Supplementary Video \href{https://drive.google.com/file/d/1nErFo7AGeqbQa2Kh2Z8HBIPJXgKWW589/view?usp=sharing}{SV1}). Thereafter, we drew N no. of samples from the underlying probability distribution (where N is the mean no. of counts across sliding windows of a given size, i.e., total number of counts across all windows divided by the number of windows for a given RP) to estimate eight common RQA variables of interest (Table 1): diagonal and vertical entropy, average diagonal \& average vertical line, maximum diagonal \& maximum vertical line, percentage of diagonal lines (determinism) and percentage of vertical lines (laminarity). We did this 1000 times to generate a bootstrap distribution of the corresponding RQA variable. From these distributions of bootstrapped samples, we computed the difference between the 95\% quantile and 5\% quantile to get the width of the range where 90\% of the data points will be distributed. Lesser the overall width of this range, higher will be the statistical confidence associated with the RQA variable estimates. For most RQA variables, this range becomes low as window size increases (Supplementary Fig. \ref{fig:S3}). We chose the window size corresponding to the knee point in these plots (i.e., the lowest window size after which increasing its value was associated with negligible change in the width of the range of the bootstrapped distribution) suggesting that the chosen window size is large enough to yield RQA variable estimates associated with high statistical confidence. For most RQA variables, this value was between 60 \& 70.  
\justifying \par We chose a common sliding window size (size = 68) across all simulated RPs to first compute the MdRQA variables for each sliding window, resulting in a distribution of these variables across sliding windows for a given RP.  To determine which summary statistic of these distributions would represent the dynamics of the system better, we computed the mean, median and mode of the distributions from the estimates from the sliding windows, which is representative of a system, and z-transformed these summary measures from sliding window distributions at the population level, where we have such estimates from multiple samples of each system. We next ran a k-nearest neighbor (KNN) classifier with nested cross validation to test which of the following aggregate measures collectively classified the time series characteristics across all RPs in the dataset with maximum cross-validation accuracy . 
\subsection{Nested Cross Validation}
\begin{figure} 
    \centering
    \includegraphics[scale=0.35]{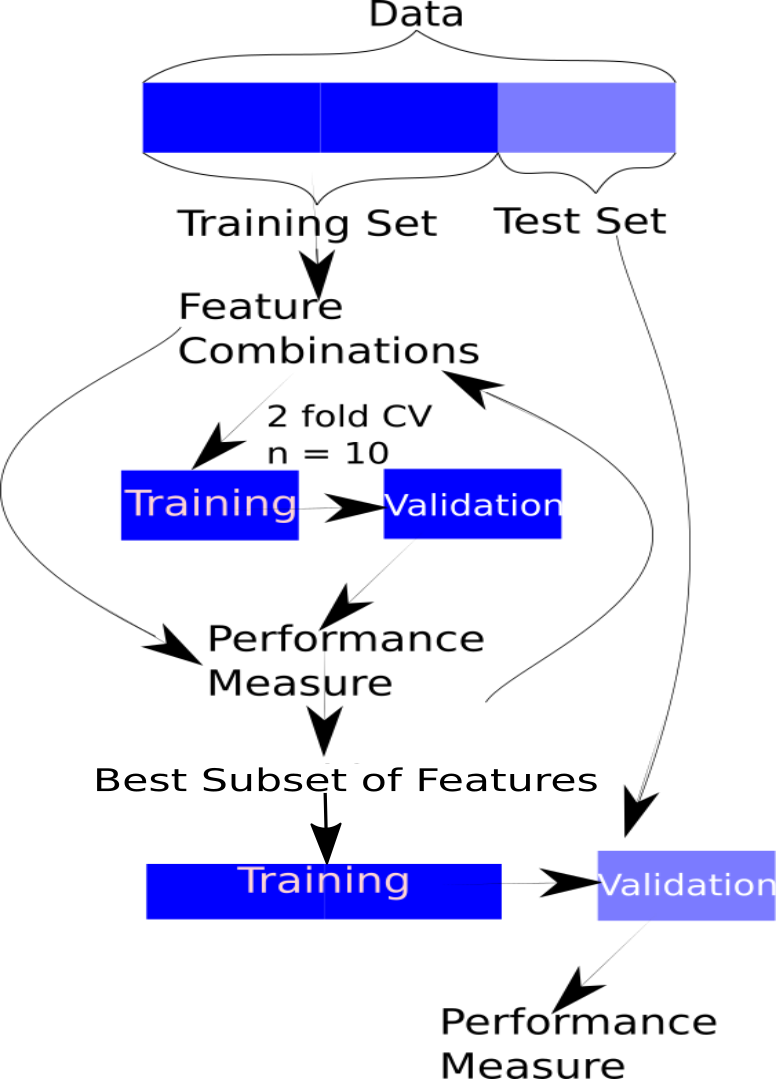}
    \caption{
    \justifying \footnotesize \textit{Schematic of nested cross validation. First branch (denoted by the arrow head in green) represents the inner loop of nested cross validation, which repeatedly divides the training set into training and validation set for the feature selection. For each combination of features, the aggregate performance score is computed to compare it with other combinations. The combination with best aggregate performance score is passed to the outer loop (denoted by arrow heads in orange), where in each iteration, the classifier is trained on the entire training dataset of that iteration, and then tested on the corresponding test set for a performance measure ( a process that is repeated over multiple iterations of the outer loop for generating a distribution of classifier performance measures).}}
    \label{fig:1}
\end{figure}
Classifier performance was quantified using a  nested cross validation approach (Fig.\ref{fig:1}) to prevent data leakage which typically occurs when test set is exposed in some form (say, for feature selection) during the training stage, likely leading to overfitting (\citeauthor{kaufman2012leakage}, \citeyear{kaufman2012leakage}, \citeauthor{verstynen2023overfitting}, \citeyear{verstynen2023overfitting}). This is particularly important when working with a smaller sample size since the number of samples in both the training as well as the test set are going to be limited which could result in either the training sample being biased or test set not being random enough. Nested cross validation overcomes this by keeping training set used for feature selection separate from that used as validation set for quantifying classifier performance. This was achieved by having an inner cross-validation loop to select features via best subset selection and an outer cross-validation loop to report classifier cross validation accuracy on the validation set corresponding to that outer loop.

For each iteration of the outer loop, data was divided into three parts. While two-thirds of it was used as a training set, one-third was kept as held out test set. Best subset selection was carried out on the training set and an aggregate performance score computed for the selected combination of features using a 2-fold repeated stratified cross validation procedure. This inner loop feature selection procedure was run for 10 iterations and the combination of features having the best performance score was then chosen and evaluated on the original held out test set of the outer loop to yield the classifier performance accuracy following a 3-fold repeated stratified cross validation procedure. The outer loop was run over 100 iterations to construct a distribution of performance scores (cross-validation accuracy and ROC) for each dataset. The performance distributions were plotted as boxplots for each measure of central tendency of MdRQA variable distributions across the sliding windows.
\subsection{Test Bed 1: Rossler Attractor}
Rossler attractor, introduced by Otto Rössler in 1976 (\citeauthor{rossler1976equation}, \citeyear{rossler1976equation}), is a simple three-dimensional dynamical system known to exhibit periodic vs. chaotic behavior depending on the parameter values a, b and c as per the following non-linear ordinary differential equations: 

\begin{equation}
    \frac{dx}{dt} = -y -z 
\end{equation}
\begin{equation}
    \frac{dy}{dt} = x + ay
\end{equation}
\begin{equation}
    \frac{dz}{dt} = b + z(x-c)
\end{equation}

\begin{figure}
    \centering
    \begin{subfigure}{0.45\textwidth}
        \includegraphics[width=\linewidth]{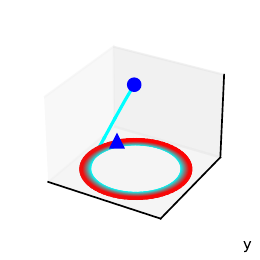}
        \caption{a=0.01}
        \label{fig:2A}
    \end{subfigure}
    \begin{subfigure}{0.45\textwidth}
        \includegraphics[width=\linewidth]{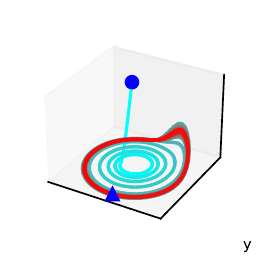}
        \caption{a=0.1}
        \label{fig:2B}
    \end{subfigure}
    \begin{subfigure}{0.45\textwidth}
        \includegraphics[width=\linewidth]{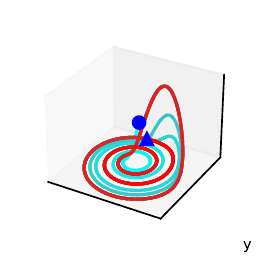}
        \caption{a=0.2}
        \label{fig:2C}
    \end{subfigure}
    \begin{subfigure}{0.45\textwidth}
        \includegraphics[width=\linewidth]{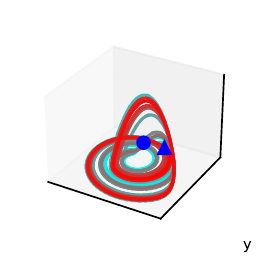}
        \caption{a=0.3}
        \label{fig:2D}
    \end{subfigure}
    \caption{\justifying \footnotesize \textit{Phase space trajectories(color gradient starting from blue and goes towards red as time goes) of Rossler attractor for different values of parameter - a. We can see a periodic behaviour when value of a is low. Blue dot is the initial state of the system and the blue triangle is final state. Blue to red gradient is across time.}}
    \label{fig:2}
\end{figure}

\begin{figure} 
    \centering
    \begin{subfigure}{0.45\textwidth}
        \includegraphics[width=\linewidth]{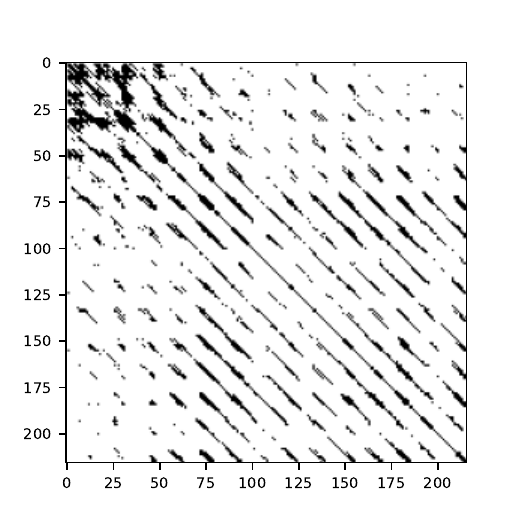}
        \caption{a=0.1}
        \label{fig:4A}
    \end{subfigure}
    \begin{subfigure}{0.45\textwidth}
        \includegraphics[width=\linewidth]{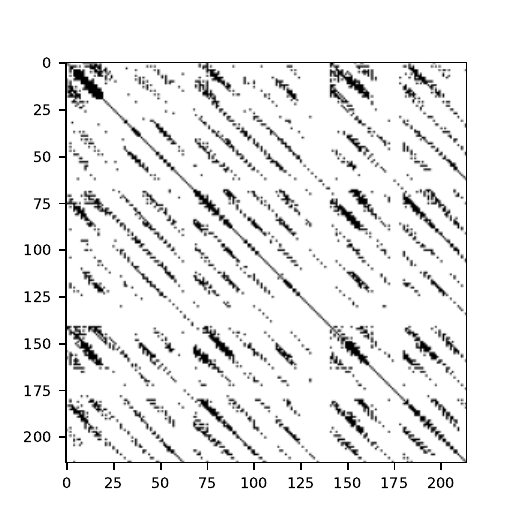}
        \caption{a=0.3}
        \label{fig:4B}
    \end{subfigure}
    \caption{\justifying \footnotesize \textit{Recurrence plot for (a) periodic and (b) chaotic Rossler attractor. }}
    \label{fig:4}
\end{figure}

Here, $\boldsymbol{x}$, $\boldsymbol{y}$, and $\boldsymbol{z}$ represent the coordinates, $\boldsymbol{t}$ represents time, and $\boldsymbol{a}$, $\boldsymbol{b}$, and c are parameters that determine the behavior of the system. The attractor is visualized by plotting the values of $\boldsymbol{x}$, $\boldsymbol{y}$, and $\boldsymbol{z}$ over time. For a given fixed value of parameters $\boldsymbol{b}$ \& $\boldsymbol{c}$  (say, as in our case, $\boldsymbol{b}$ = 0.2, and $\boldsymbol{c}$ = 5.7) and a range of values of a, the attractor is known to show periodic behavior (for values of $\boldsymbol{a}$ = 0.01 to <0.2; Fig. 2), and chaotic behaviour (for values of $\boldsymbol{a}$ = 0.2 to 0.4; Fig. 2) (Figures \ref{fig:2} \& \ref{fig:3}). Figure \ref{fig:4} shows the RPs for periodic and chaotic attractors.

\subsubsection{Validation Approach 1: predicting periodic vs. chaotic state of Rossler Attractor }
We simulated the attractors under nine different levels of noise conditions (SNR or signal to noise ratio values of 0.125, 0.25, 0.5, 0.75, 1.0, 1.25, 1.5, 1.75, 2.0), to test the sensitivity of classifier performance measure against gaussian noise. The choice of SNRs was arbitrary, but with the aim of including a high noise region as an extreme case. We used five values of \textbf{a} (0.1, 0.15, 0.2, 0.25, 0.3), and simulated 10 RPs for each combination of \textbf{a} and SNR with different initial values drawn from a random normal distribution. This added up to a total of 50 RPs per noise level and 450 RPs in all. The lengths of time series were randomly sampled from a uniform distribution of lengths 250-450 samples, sampled at the rate of 0.25Hz (corresponding to 1000-1800 seconds, with $dt = 10^{-4}$). Whole RPs were generated using a fixed recurrence rate (~10\%). We used the sliding window technique described earlier to obtain the distributions of nine MdRQA variables (recurrence rate, average diagonal line length, average vertical line length, percentage laminarity, percentage determinism, vertical entropy, diagonal entropy, vertical maximum line length, diagonal maximum line length; for more details see Table \ref{table-1}) across all sliding windows of each RP. The central tendencies (mean, median and mode) of the MdRQA variable distributions were run through the nested cross validation method described above separately for each noise level.

\subsubsection{Results of Validation Approach 1}
\begin{figure} 
    \centering
    \includegraphics[width=\linewidth]{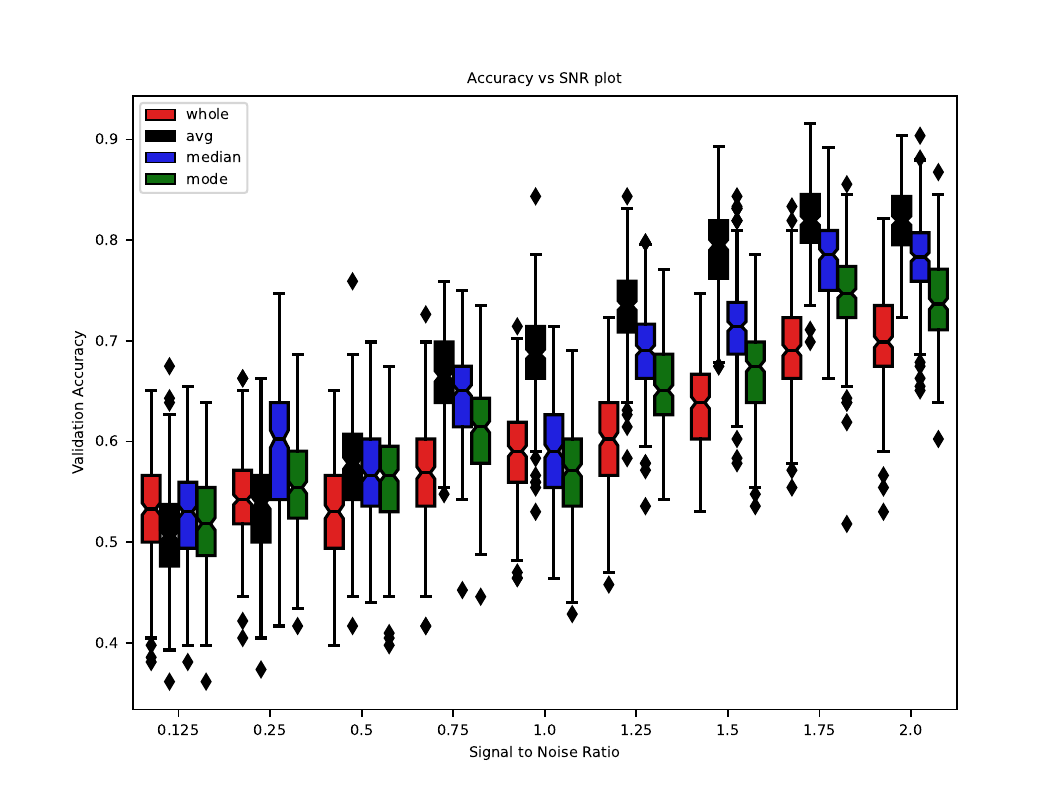}
    \caption{\justifying \footnotesize \textit{Box plots of classifier accuracy estimates from the validation set in the nested cross validation using average, median, mode of MdRQA variable distributions across sliding windows as well as MdRQA variables of the whole original RP }}
    \label{fig:5}
\end{figure}
\begin{figure} 
    \centering
    \includegraphics[width=\linewidth]{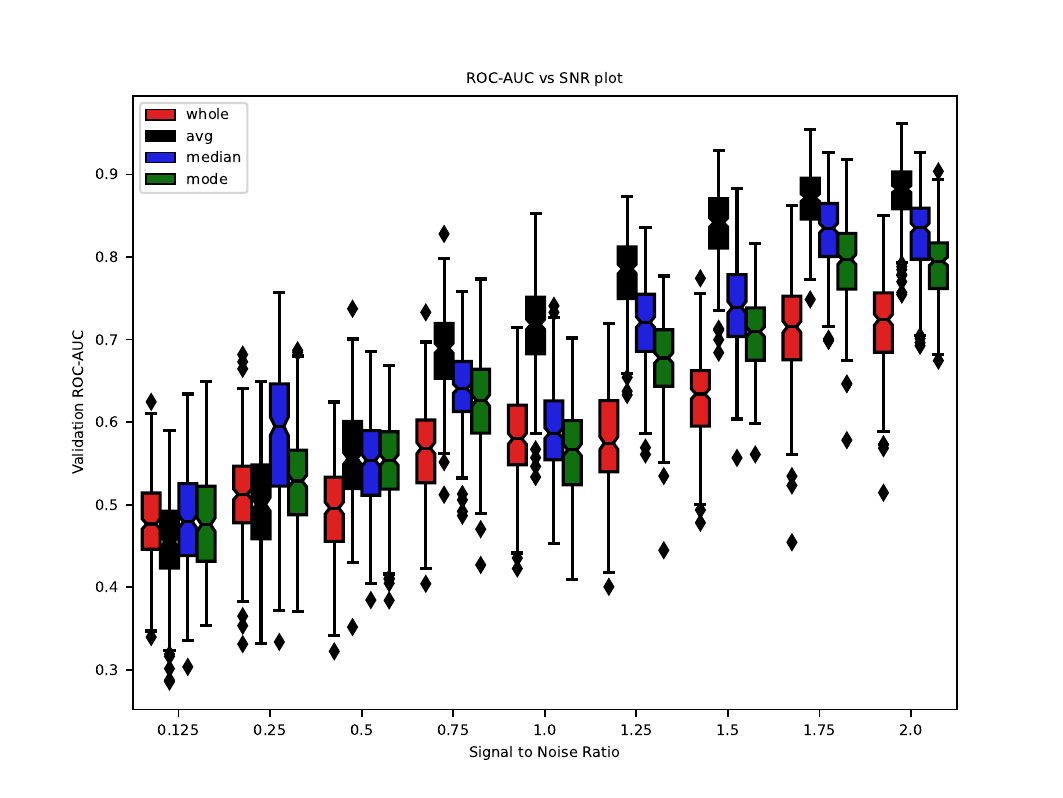}
    \caption{\justifying \footnotesize \textit{Box plots of classifier ROC AUC estimates from the validation set in the nested cross validation using average, median, mode of MdRQA variable distributions across sliding windows as well as MdRQA variables of the whole original RP }}
    \label{fig:6}
\end{figure}
Figures \ref{fig:5} and \ref{fig:6} show the box plots for classifier cross validation accuracy and ROC AUC measures. We found that under very high noise levels ($SNR \leq 0.25$), the classifiers discriminating between periodic vs. chaotic state of the Rossler attractor using mean, median and mode values of the MdRQA variable distributions across sliding windows performed at chance in terms of accuracy ($\text{mean CV accuracy}_{mean,SNR\leq0.25}$ = 0.53, CI=[0.435, 0.624], $\text{mean CV accuracy}_{median,SNR\leq0.25}$=0.526, CI=[0.427, 0.624], $\text{mean CV accuracy}_{mode,SNR\leq0.25}$=0.542, CI = [0.439, 0.645]) as well as ROC AUC ($\text{mean CV ROC-AUC}_{mean,SNR\leq0.25}$ = 0.497, CI=[0.382, 0.672], $\text{mean CV ROC-AUC}_{median,SNR\leq0.25}$ = 0.494, CI=[0.377, 0.612], $\text{mean CV ROC-AUC}_{mode,SNR\leq0.25}$ = 0.516, CI = [0.389, 642]). Interestingly, under same noise levels,  the classification based on MdRQA variable from the whole RP (not using the sliding window technique) also performed at chance levels ($\text{mean CV ROC-AUC}_{whole RP,SNR\leq0.25}$ = 0.507, CI = [0.378, 0.637], $\text{mean CV accuracy}_{mean,SNR\leq0.25}$ = 0.540, CI=[0.434, 0.647] ). However, as the strength of signal increased, a steady increase in classifier performance was found across the three summary measures as well as the MdRQA variables from the whole RP, even though the latter was found to fare worse than the former, in general. To quantify this overall trend across SNRs, 
\begin{figure} 
    \centering
    \includegraphics[width=\linewidth]{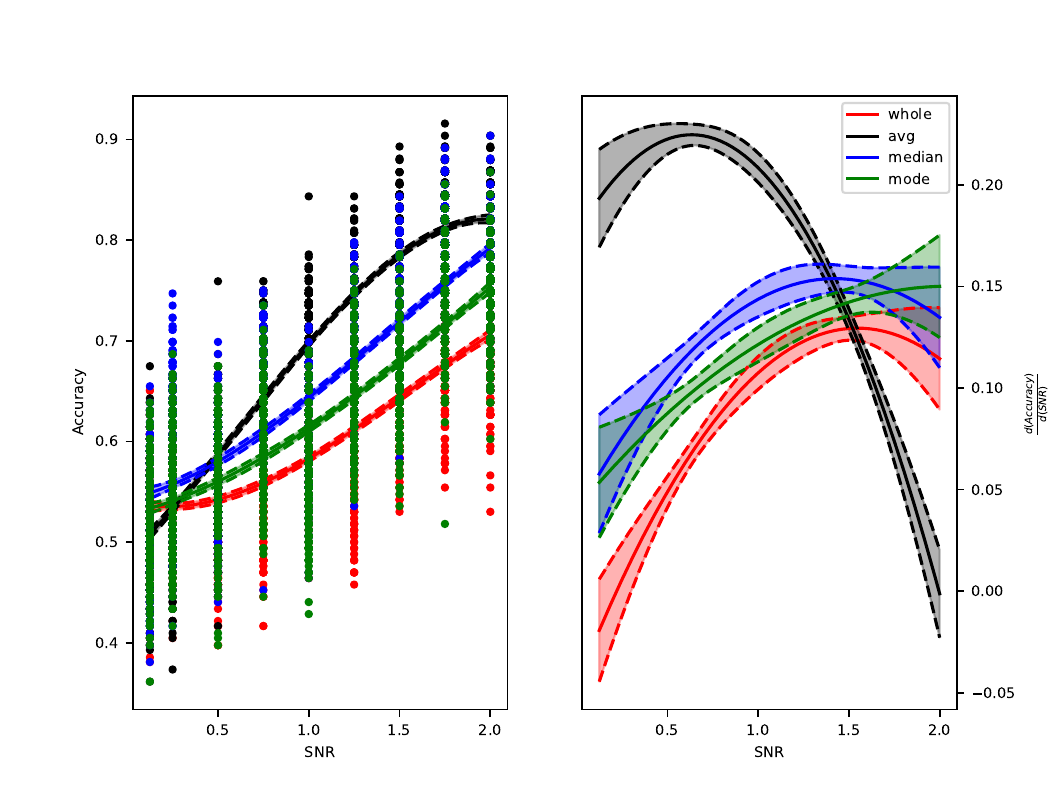}
    \caption{\justifying \footnotesize \textit{Spline fit (left) and local slopes (first derivative- right) of the accuracy-SNR data. Observed that the whole RP underperforms for the entire range of SNR tested. Mean estimate from sliding windows perform badly under high noise condition (SNR<0.25), but the faster variation with SNR makes it possible to perform better than other estimates. Mode and median shows a diverging trend, where accuracy scores for median increases quickly for later points. }}
    \label{fig:7}
\end{figure}
\begin{figure} 
    \centering
    \includegraphics[width=\linewidth]{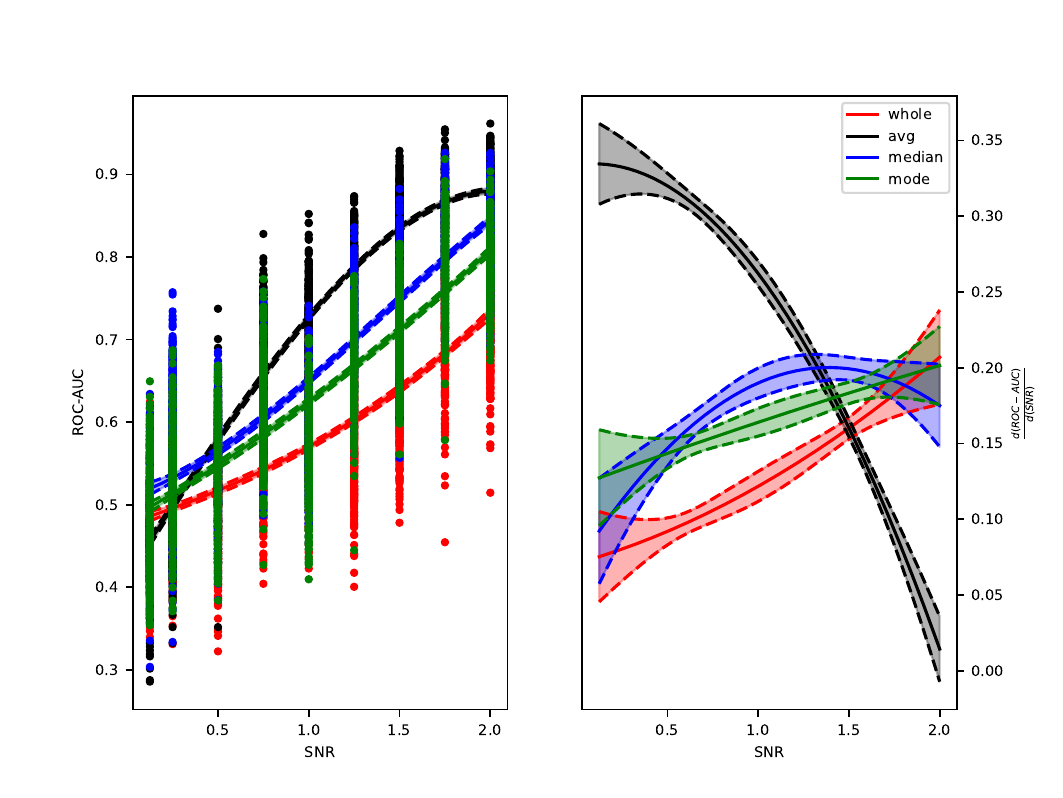}
    \caption{\justifying \footnotesize \textit{Spline fit (left) and local slopes (first derivative- right) of the ROC AUC-SNR data. Measures from the whole RP underperformed for the entire range of SNR tested. Mean estimate from sliding windows performed badly under high noise condition (SNR<0.25), but with increasing SNRs, performed better than other estimates. Mode and median shows a diverging trend, where accuracy scores for median increases quickly for later points. }}
    \label{fig:8}
\end{figure}

\begin{table} 
\begin{tabular}{ |p{3cm}||p{3cm}|p{3cm}|p{3cm}|  }
 \hline
 \multicolumn{4}{|c|}{Table 2} \\
 \hline
 Estimate & Slope(m) & 2.5\% quantile & 97.5\% quantile \\
 \hline
 Whole RP & 0.094 & 0.091 & 0.097 \\
 \hline 
 mean & 0.178 & 0.173 & 0.181 \\
 \hline 
 median & 0.133 & 0.128 & 0.136 \\
 \hline 
 mode & 0.118 & 0.114 & 0.121 \\
 \hline
\end{tabular}
\caption{\justifying \footnotesize \textit{Global slope of accuracy-SNR data. It was found that the mode has the least slope among all central tendency measures estimated, suggesting its low sensitivity to noise. }}
\label{table-2}
\end{table}

\begin{table} 
\begin{tabular}{ |p{3cm}||p{3cm}|p{3cm}|p{3cm}|  }
 \hline
 \multicolumn{4}{|c|}{Table 3} \\
 \hline
 Estimate & Slope(m) & 2.5\% quantile & 97.5\% quantile \\
 \hline
 Whole RP & 0.128 & 0.124 & 0.132 \\
 \hline 
 mean & 0.236 & 0.23 & 0.24 \\
 \hline 
 median & 0.177 & 0.171 & 0.181 \\
 \hline 
 mode & 0.165 & 0.160 & 0.168 \\
 \hline
\end{tabular}
\caption{\justifying \footnotesize \textit{Global slope of ROC AUC-SNR data. It was found that the mode has the least slope among all central tendency measures estimated, suggesting its low sensitivity to noise. }}
\label{table-3}
\end{table}

we fit a spline regression curve to the SNR-accuracy/ROC AUC data and estimated the first derivative of these curves locally to plot the change in performance measure in each of the finite interval (Figures \ref{fig:7} \& \ref{fig:8}).  We also estimated the slope of the entire distribution by fitting a line and the region where the slopes would be distributed for 95\% of the time, by computing the 2.5\% quantile and 97.5\% quantile of the slope distribution resulting from bootstrapping (1000 iterations, by doing sampling with replacement). We found the slope of classifier performance across noise levels to be the least for the mode of the MdRQA variable distributions across sliding windows (see table \ref{table-2} \& \ref{table-3}) indicating it to be the most robust in discriminating the different dynamic states of the system across varying noise levels.  On the other hand, the performance measures based on the mean of MdRQA variable distribution were most sensitive to noise levels in the data. 
\subsection{Test Bed 2: Kuramoto model}
\begin{figure} 
    \centering
    \includegraphics[width=\linewidth]{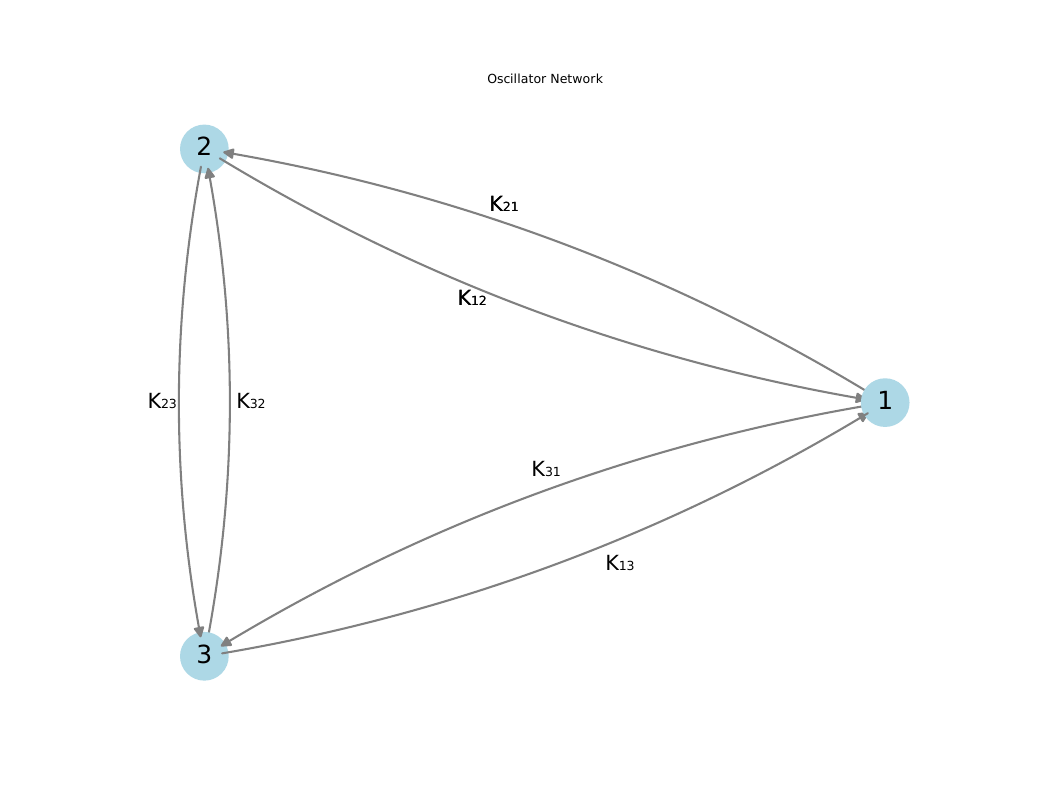}
    \caption{\justifying \footnotesize \textit{Pictorial depiction of a Kuramoto model with three oscillators. Here Kij are the edges from jth oscillator to ith oscillator.}}
    \label{fig:9}
\end{figure}
The Kuramoto system of coupled oscillators(\citeauthor{kuramoto1975international}, \citeyear{kuramoto1975international}; \citeauthor{strogatz2000kuramoto}, \citeyear{strogatz2000kuramoto}), famously used to model the synchronous flashing of light in populations of fireflies (\citeauthor{ermentrout1991adaptive},\citeyear{ermentrout1991adaptive}), is defined to consist of a population of N coupled phase oscillators each with a natural frequency $\omega_{i}$ sampled from a distribution $g(\omega)$. Figure \ref{fig:9} shows a schematic for a system of three coupled oscillators. At t=0, each oscillator has a phase value $\theta_{i}(0)$, which is updated over time according to partial differential equation given below: 
\begin{equation}
    \dot{\theta_{i}} = \omega_{i} + \sum_{j=1}^{N} K_{ij} \sin{(\theta_{j}-\theta{i})}, i=1, ..., N
\end{equation}
where $K_{ij}$ is the coupling strength for the edge (or the connection between) from $j^{th}$ oscillator to the $i^{th}$ oscillator. Critical coupling strength, $K_{c}$, is defined as the difference between the maximum and minimum values of natural frequencies across oscillators in the system.
\begin{equation}
    K_{c} = |\omega_{max}-\omega_{min}|
\end{equation}
 In the simplest case scenario that we considered here, known as mean field coupling, the coupling strength, $K_{ij}$, for all edges (or between any two oscillators) in the network of oscillators was assumed to be the same. $K_{ij}$ under these conditions is given by:

\begin{equation}
    K_{ij} = K/N >0, \forall i,j \in \{1,2,3,...,N\}
\end{equation}

The synchrony of the system is defined in terms of a complex value order parameter, r, obtained in our case as follows: 
\begin{equation}
    r e^{i\psi} =  \frac{1}{N} \sum_{j=1}^{N}e^{i \theta_{j}}
\end{equation}

Here $\psi$ is the average phase value. To arrive at  an expression that makes the dependence of synchrony on values of K explicit, we begin with multiplying both sides by $e^{-i \theta_{i}}$.
\begin{equation}
    r e^{i\psi} e^{-i \theta_{i}} = \left( \frac{1}{N} \sum_{j=1}^{N}e^{i \theta_{j}}\right) e^{-i \theta_{i}}
\end{equation}
\begin{equation}
    r e^{i(\psi -\theta_{i})}= \frac{1}{N} \sum_{j=1}^{N}e^{i(\theta_{j}-\theta_{i})}
\end{equation}
\begin{equation}
    r \sin{(\psi -\theta_{i})} = \frac{1}{N} \sum_{j=1}^{N}\sin{(\theta_{j}-\theta_{i})}
\end{equation}
One may note the similarity of the right hand side of the above  equation with that  of equation (11), except that the latter  was multiplied by a factor of K. Using the left hand side of the above equation(15),  equation (11) may be rewritten as follows:
\begin{equation}
    \dot{\theta_{i}} = \omega_{i} + K r \sin{(\psi -\theta_{i})}
\end{equation}
Equation 15 suggests that as $K$, tends to zero (low coupling strength), $K r \sin{(\psi -\theta_{i})}$would become negligibly small compared to $\omega_{i}$ resulting in the following approximation:
\begin{equation}
    \dot{\theta_{i}} \approx \omega_{i}
\end{equation}
\begin{equation}
    \theta_{i}(t) \approx \omega_{i} t + \theta_{i}(0)
\end{equation}
which suggests that  the oscillator would simply oscillate at  its natural frequency or in other words, the oscillations of other oscillators would have no influence on  it. On the other hand, at higher values of $K$ or when the coupling strength is tending to infinity, as per equation 15, the natural frequency would become negligible when compared to the $K r \sin{(\psi -\theta_{i})}$ term and consequently, the oscillator’s phase value would converge to the average phase value of the population. 

\begin{figure}[!t]
  \begin{tabular}[b]{ p{4.5cm}p{4.5cm}}
    \begin{tabular}[b]{c}
      \begin{subfigure}[b]{0.35\columnwidth}
        \includegraphics[width=0.8\textwidth]{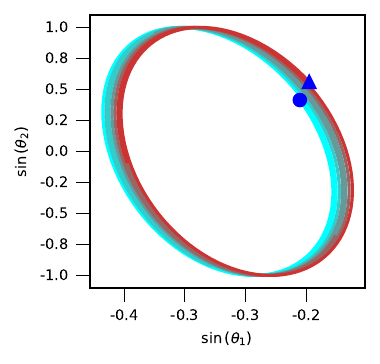}
        \caption{}
        \label{fig:10A}
      \end{subfigure}\\
      \begin{subfigure}[b]{0.35\columnwidth}
        \includegraphics[width=0.8\textwidth]{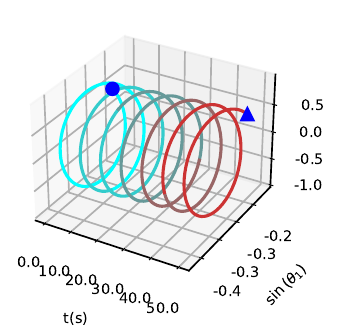}
        \caption{}
        \label{fig:10B}
      \end{subfigure}
    \end{tabular}
    &
    \begin{subfigure}[b]{0.35\columnwidth}
      \includegraphics[width=1.5\textwidth]{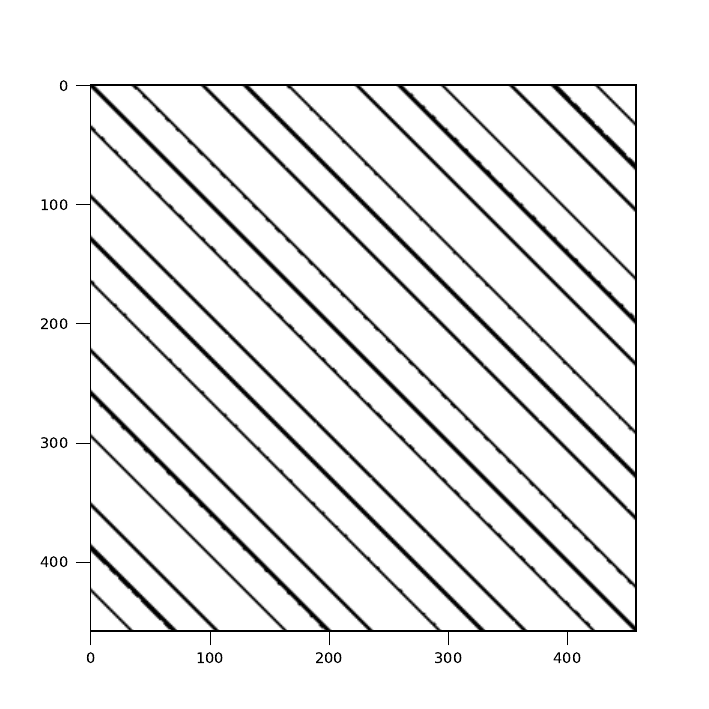}
      \caption{}
      \label{fig:10C}
    \end{subfigure}
  \end{tabular}
  \caption{\justifying \footnotesize \textit{Simulation from Kuramoto model with mean field coupling for two oscillators, where coupling strength is one tenth of the critical coupling strength. We can see that the trajectories go almost in circles, resulting in parallel lines in the RP. This is also expected from Poincare’s recurrence theorem, when the number of oscillators is few. However, in low coupling strength ($<K_{c}$), the trajectories need not always be circular like the one we have; they could also exhibit complicated shapes, still showing repetitions over time. Such trajectories can temporarily have synchrony as well.}}
  \label{fig:10}
\end{figure}

\begin{figure}[!t]
  \begin{tabular}[b]{ p{4.5cm}p{4.5cm}}
    \begin{tabular}[b]{c}
      \begin{subfigure}[b]{0.35\columnwidth}
        \includegraphics[width=0.8\textwidth]{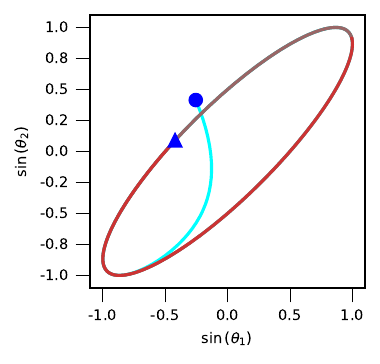}
        \caption{}
        \label{fig:11A}
      \end{subfigure}\\
      \begin{subfigure}[b]{0.35\columnwidth}
        \includegraphics[width=0.8\textwidth]{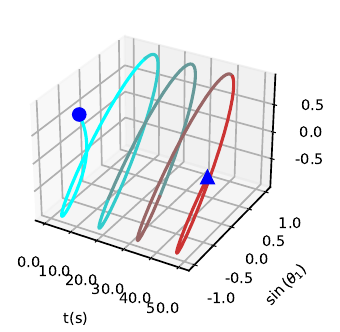}
        \caption{}
        \label{fig:11B}
      \end{subfigure}
    \end{tabular}
    &
    \begin{subfigure}[b]{0.35\columnwidth}
      \includegraphics[width=1.5\textwidth]{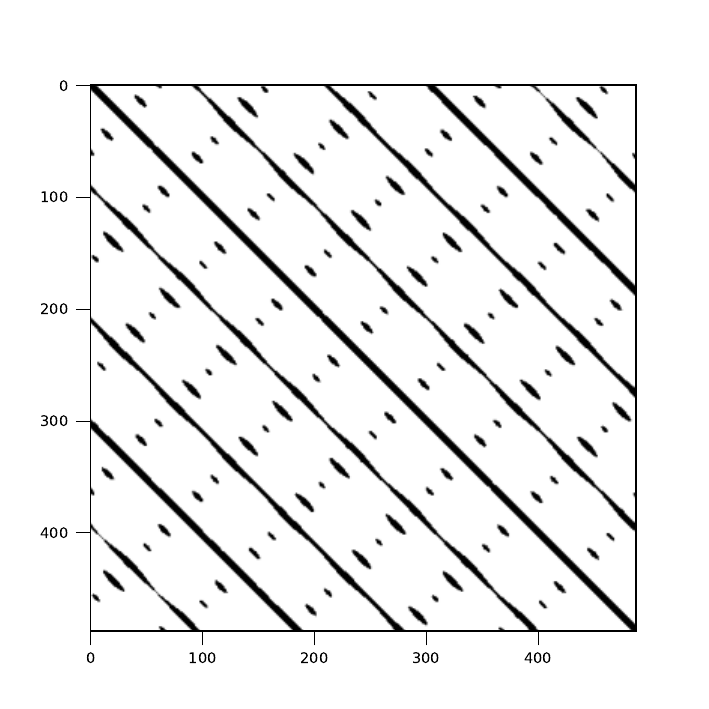}
      \caption{}
      \label{fig:11C}
    \end{subfigure}
  \end{tabular}
  \caption{\justifying \footnotesize \textit{Simulation from the Kuramoto model with mean field coupling for two oscillators, where the coupling strength is equal to the critical coupling strength. We can see that the trajectories go almost in an ellipse, where the major axis is much larger than the minor axis, resulting in parallel lines in the recurrence plot (RP) with interconnected structures that are also parallel. This pattern aligns with the expectations of Poincare's recurrence theorem when the number of oscillators is small.}}
  \label{fig:11}
\end{figure}

\begin{figure}[!t]
  \begin{tabular}[b]{ p{4.5cm}p{4.5cm}}
    \begin{tabular}[b]{c}
      \begin{subfigure}[b]{0.35\columnwidth}
        \includegraphics[width=0.8\textwidth]{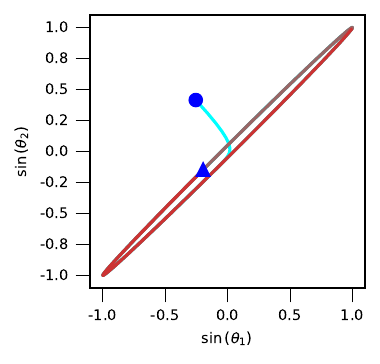}
        \caption{}
        \label{fig:12A}
      \end{subfigure}\\
      \begin{subfigure}[b]{0.35\columnwidth}
        \includegraphics[width=0.8\textwidth]{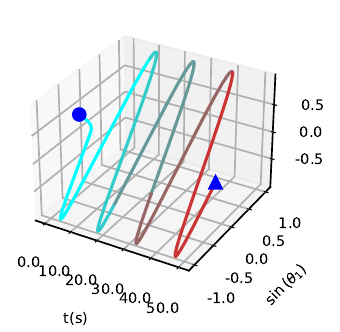}
        \caption{}
        \label{fig:12B}
      \end{subfigure}
    \end{tabular}
    &
    \begin{subfigure}[b]{0.35\columnwidth}
      \includegraphics[width=1.5\textwidth]{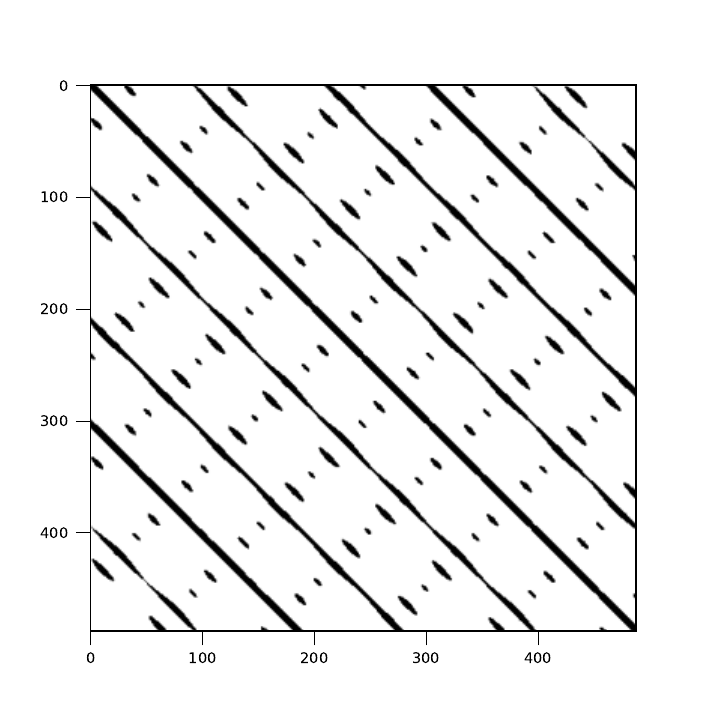}
      \caption{}
      \label{fig:12C}
    \end{subfigure}
  \end{tabular}
  \caption{\justifying \footnotesize \textit{Simulation from the Kuramoto model with mean field coupling for two oscillators, where the coupling strength is ten times the critical coupling strength. We can see that the trajectories go almost in an ellipse, approaching a line, where the major axis is much larger than the minor axis, resulting in parallel lines in the recurrence plot (RP) with interconnected structures that are also parallel. This pattern aligns with the expectations of Poincare's recurrence theorem when the number of oscillators is small.}}
  \label{fig:12}
\end{figure}

\begin{figure}[!t]
  \begin{tabular}[b]{ p{4.5cm}p{4.5cm}}
    \begin{tabular}[b]{c}
      \begin{subfigure}[b]{0.35\columnwidth}
        \includegraphics[width=0.8\textwidth]{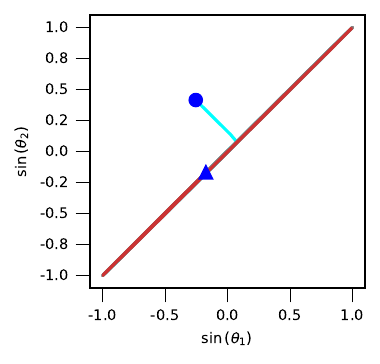}
        \caption{}
        \label{fig:13A}
      \end{subfigure}\\
      \begin{subfigure}[b]{0.35\columnwidth}
        \includegraphics[width=0.8\textwidth]{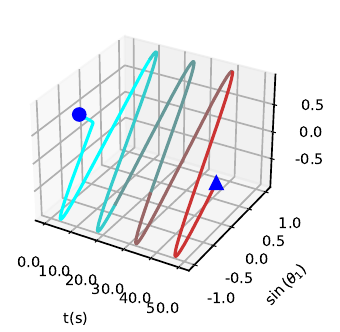}
        \caption{}
        \label{fig:13B}
      \end{subfigure}
    \end{tabular}
    &
    \begin{subfigure}[b]{0.35\columnwidth}
      \includegraphics[width=1.5\textwidth]{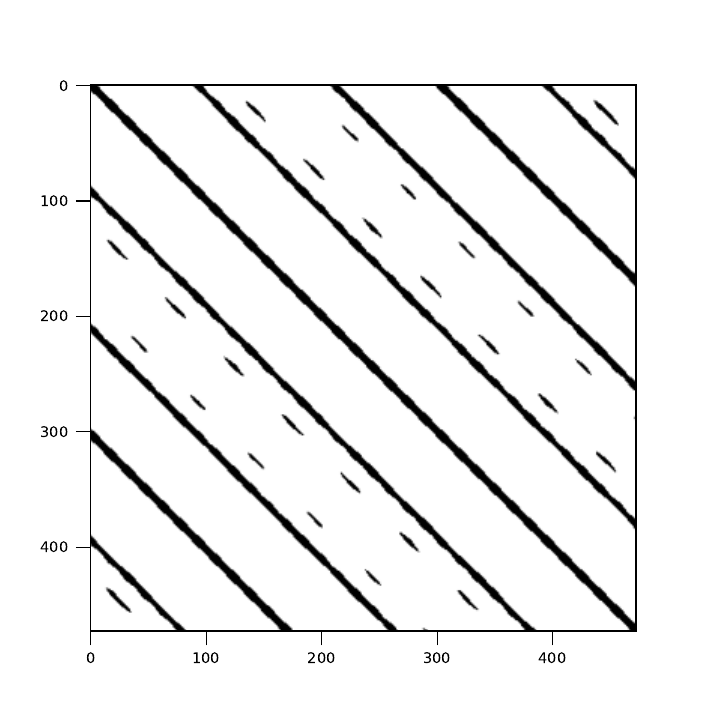}
      \caption{}
      \label{fig:13C}
    \end{subfigure}
  \end{tabular}
  \caption{\justifying \footnotesize \textit{Simulation from the Kuramoto model with mean field coupling for two oscillators, where the coupling strength is hundred times the critical coupling strength. We can see that the trajectories go almost in an ellipse, approaching a line, where the major axis is much larger than the minor axis, resulting in parallel lines in the recurrence plot (RP) with interconnected structures that are also parallel. This pattern aligns with the expectations of Poincare's recurrence theorem when the number of oscillators is small.}}
  \label{fig:13}
\end{figure}

\begin{equation}
    K_{c} = |\omega_{max} - \omega_{min}|
\end{equation}
\justifying \par The implication of above derivation on the phase space and recurrence plot can be illustrated using a simple scenario, say, a system of two oscillators while we examine its dynamics for three coupling strengths, say, 0.1, 1 and 10 times the critical coupling strength of the system, $K_{c}$(\citeauthor{biccari2020stochastic}, \citeyear{biccari2020stochastic}). Figure \ref{fig:10} panel A plots the case of $K = 0.1 K_{c}$ where the trajectories could be seen to revisit the phase space cyclically. Fig \ref{fig:10} shows how this cyclical trajectory translates to a RP with multiple symmetrically spaced, parallel diagonal lines.  However, unlike the Rossler attractor, which would either show a periodic or chaotic behaviour, the Kuramoto system with two oscillators  always shows periodic behaviour at coupling strength lesser than the critical coupling strength (Fig. \ref{fig:10}; \citeauthor{acebron2005kuramoto}, \citeyear{acebron2005kuramoto}; \citeauthor{chopra2009exponential}, \citeyear{chopra2009exponential}; \citeauthor{biccari2020stochastic}, \citeyear{biccari2020stochastic}). 

At critical coupling strength ($K = K_{c}$, Fig. \ref{fig:11}), the eccentricity of the trajectory in the phase space  can be found to increase and take the form of an ellipse. In such a scenario, some points in the trajectory  would be closer  than others, resulting in the RP a  pattern of lines that are somewhat broken but periodic and parallel to the main diagonal  (Fig. \ref{fig:11}, panel C). 

At a coupling strength greater than the critical coupling strength (say, $K = 10K_{c}$, Figure \ref{fig:12}), while the RP continues to look similar, the trajectory takes the shape of a narrower ellipse. As  K tends to infinity (or a very high value) the trajectory looks almost like a line and the oscillators  are said to be phase locked or in sync. The resulting RP is shown in Fig. \ref{fig:13}.
\begin{figure} 
    \centering
    \includegraphics[width=0.5\textwidth]{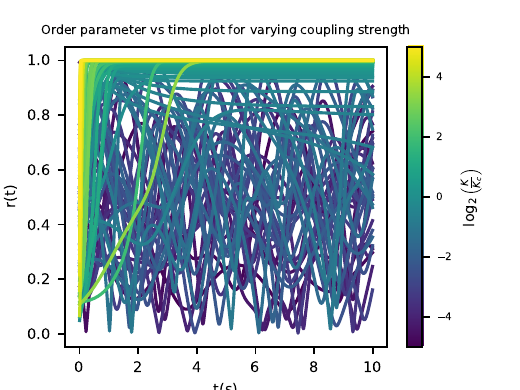}
    \caption{\justifying \footnotesize \textit{For a system of four oscillator, evolution of synchrony over time, for varying amount of coupling strength. The color code is for the log of ratio between the coupling strength and critical coupling strength(base 2). We can see that, when the log ratio is less than zero, even if the system reaches synchrony temporarily, it was unable to sustain it. But, for the log ratio grater than zero, after reaching synchrony, it continues to be in that state. }}
    \label{fig:14}
\end{figure}
To graphically illustrate how the synchrony of this two oscillator Kuramoto system (given by the order parameter as per Equation 12) is related to the coupling strength of its oscillators, we varied the latter as a function of critical coupling strength by multiplying it with powers of two (100 values at regular intervals, from $2^{-5}$ to $2^{5}$, denoted by the color bar) and plotted the resulting temporal evolution of the order variable, where the log is having 2 as its base. As can be seen in Fig. \ref{fig:14}, while the order parameter fluctuates more at lower coupling strengths, they seem to saturate at higher coupling strengths, consistent with equation 16 and the discussion following it.
\begin{figure}[!t]
    \begin{tabular}[b]{ p{2.5cm}p{2.5cm}p{2.5cm}p{2.5cm}p{2.5cm}}
        \begin{subfigure}[b]{0.35\columnwidth}
            \includegraphics[width=0.5\textwidth]{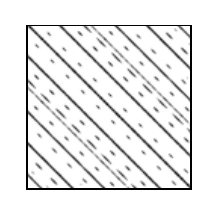}
            \caption{ }
            \label{fig:15A}
        \end{subfigure}
        &
        \begin{subfigure}[b]{0.35\columnwidth}
            \includegraphics[width=0.5\textwidth]{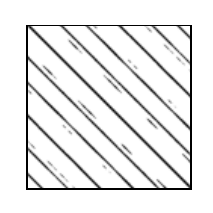}
            \caption{ }
            \label{fig:15B}
        \end{subfigure}
        &
        \begin{subfigure}[b]{0.35\columnwidth}
            \includegraphics[width=0.5\textwidth]{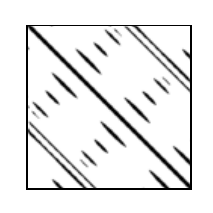}
            \caption{ }
            \label{fig:15C}
        \end{subfigure}
        &
        \begin{subfigure}[b]{0.35\columnwidth}
            \includegraphics[width=0.5\textwidth]{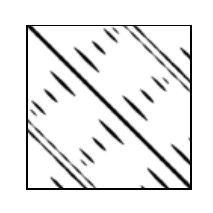}
            \caption{ }
            \label{fig:15D}
        \end{subfigure}
        &
        \begin{subfigure}[b]{0.35\columnwidth}
            \includegraphics[width=0.5\textwidth]{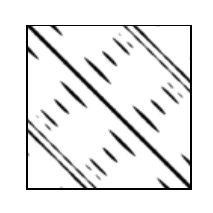}
            \caption{ }
            \label{fig:15E}
        \end{subfigure}
    \end{tabular}
    \caption{\justifying \footnotesize \textit{Recurrence plots for N=3, $K$= 0.01 (a), 0.1 (b), 1 (c), 10 (d), 100 (e) times $K_c$}}
    \label{fig:15}
\end{figure}

\justifying \par As we increase the number of oscillators to 3   and varied the coupling strength as a function of critical coupling strength (by multiplying it with powers of 10 from $10^{-2}$ to $10^{2}$), we find that while at low coupling strengths, the RP is constituted by solid or broken lines parallel to the main diagonal (Fig. \ref{fig:15}), at coupling strengths greater than the critical coupling strength, the broken lines parallel to the diagonal lines become more prominent. These results are consistent with the proposed idea of using recurrence plots to distinguish between a dynamical system having coupling strength lesser vs. greater than the critical coupling strength.
\subsubsection{Validation Approach 2: Predicting whether the coupling strength is greater than that of the critical coupling strength in the mean-field model of the Kuramoto system}
\justifying \par We used the Kuramoto system of coupled oscillators to test the ability of the sliding window based approach in predicting whether the system is strongly coupled ($K>K_{c}$) or not ($K\leq K_{c}$). The lengths of time series were randomly sampled from a uniform distribution of lengths 150-450 samples, sampled at the rate of 10Hz (corresponding to 15-45 seconds, with $dt = 10^{-2}$). In addition, the number of oscillators was randomly sampled from a range of 3 to 6. Coupling strength was randomly sampled from a uniform distribution $[0, 2K_{c}]$, such that the probability of picking a coupling strength less than $K_{c}$ is equal to that of picking a coupling strength greater than $K_{c}$. To test how our approach fares across a range of gaussian noise added to the signal, we simulated data for nine values of signal to noise ratios (0.125, 0.5, 1.0, 1.5, 2.0, 2.5, 3.0, 3.5, 4.0). Then, as in case of the earlier example of Rossler Attractors, we constructed RPs and computed the summary statistics of the MdRQA variables across sliding windows, z-transformed them before running the nested cross validation procedure for classifier performance.
\subsubsection{Results of Validation Approach 2}
\begin{figure} 
    \centering
    \includegraphics[width=\textwidth]{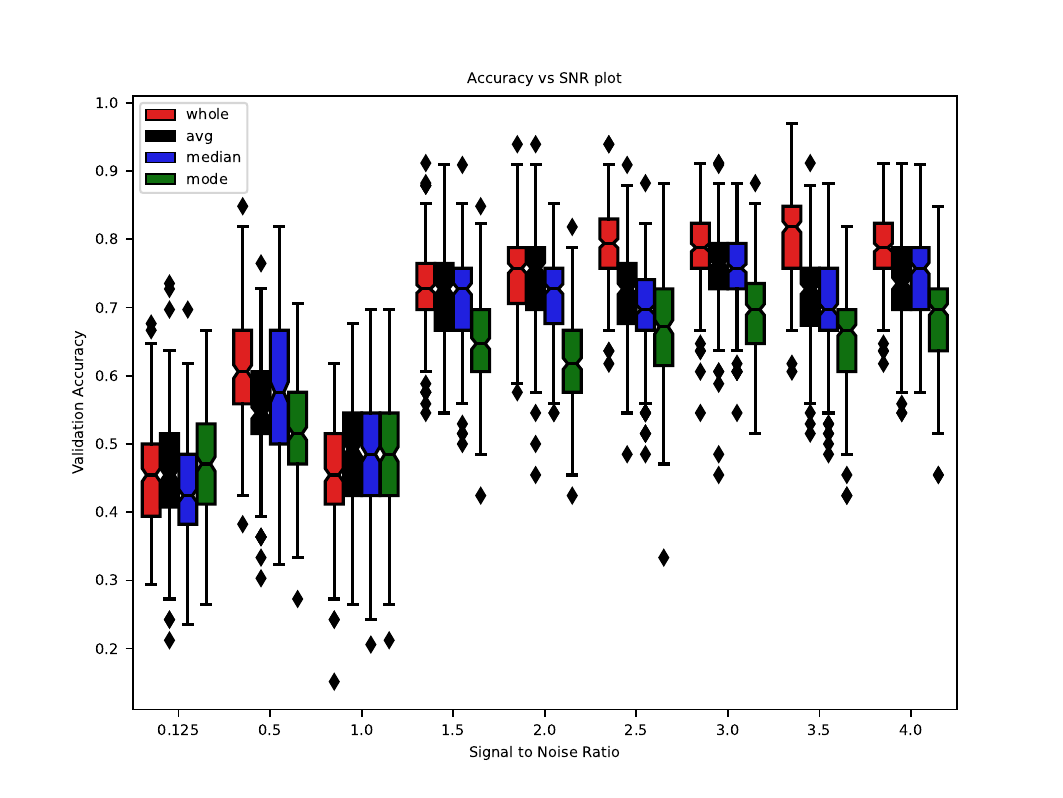}
    \caption{\justifying \footnotesize \textit{Box plot of classifier accuracy estimates on validation sets from nested cross validation using average, median, mode of MdRQA variable distributions across sliding windows as well as MdRQA variables of the whole original RP. }}
    \label{fig:16}
\end{figure}
\begin{figure} 
    \centering
    \includegraphics[width=\textwidth]{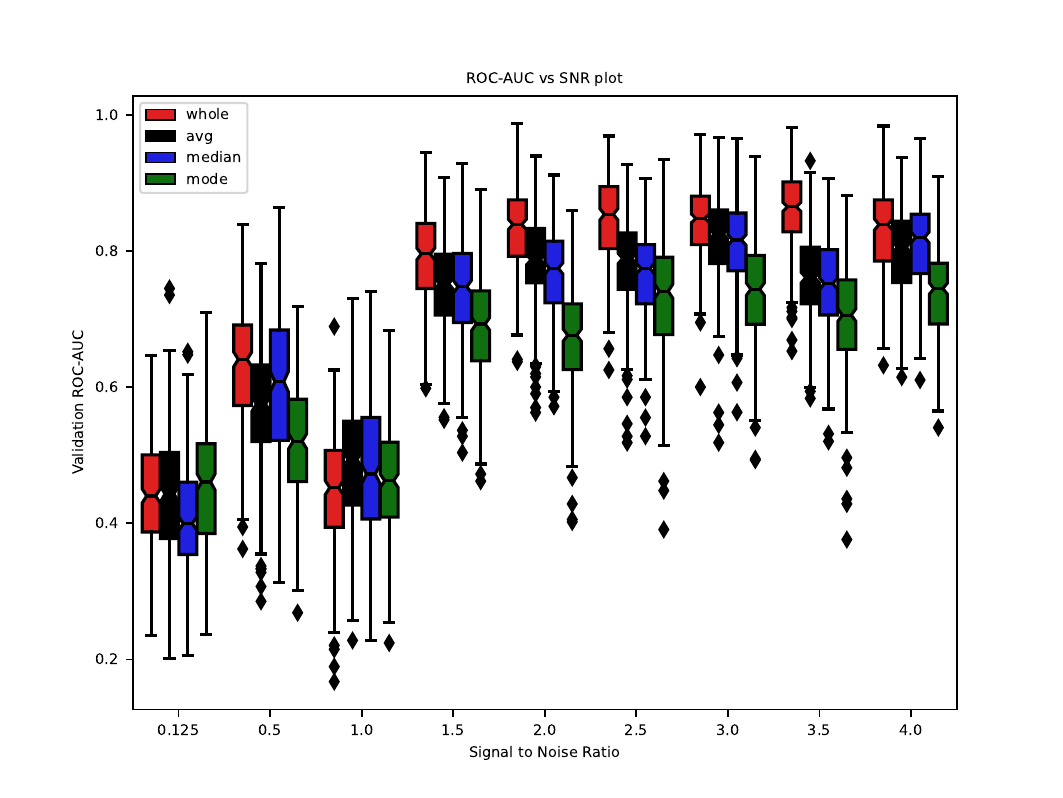}
    \caption{\justifying \footnotesize \textit{Box plot of classifier ROC AUC estimates on validation sets from nested cross validation using average, median, mode of MdRQA variable distributions across sliding windows as well as MdRQA variables of the whole original RP.}}
    \label{fig:17}
\end{figure}
\begin{figure} 
    \centering
    \includegraphics[width=\textwidth]{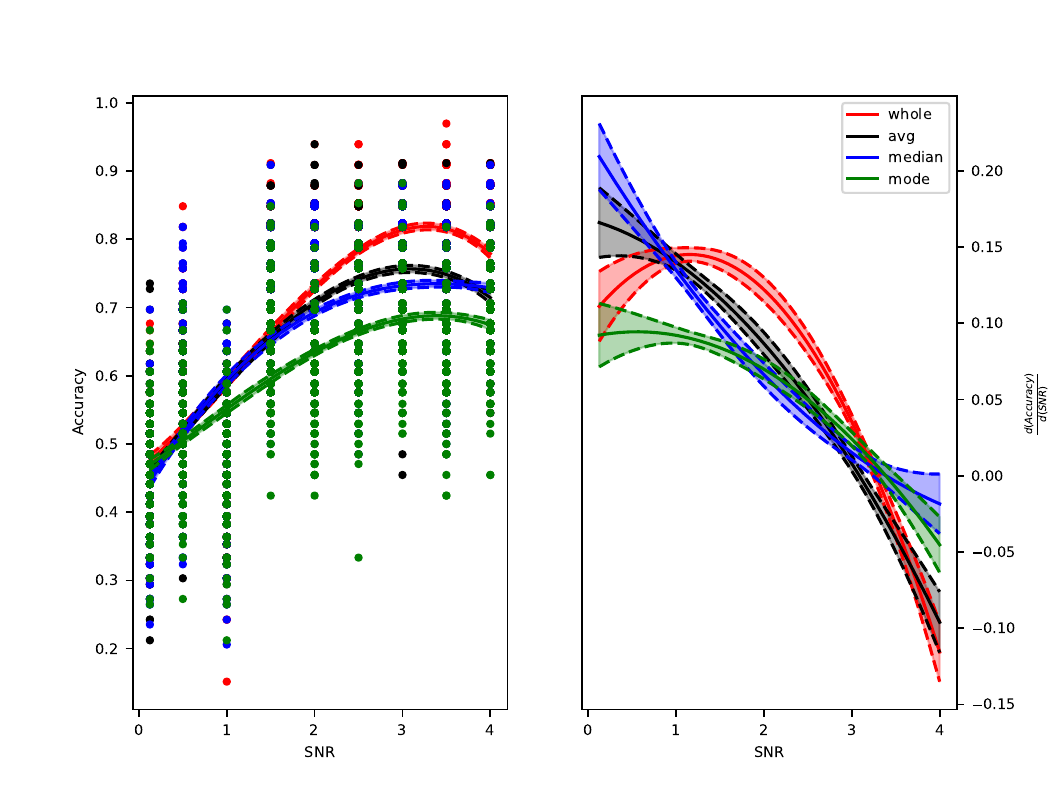}
    \caption{\justifying \footnotesize \textit{Spline fit (left) and local slopes (first derivative- right) of the accuracy-SNR data. While at high noise condition ($SNR\leq0.5$) all measures performed similarly at chance level, performance improved with increasing SNRs.}}
    \label{fig:18}
\end{figure}
\begin{figure} 
    \centering
    \includegraphics[width=\textwidth]{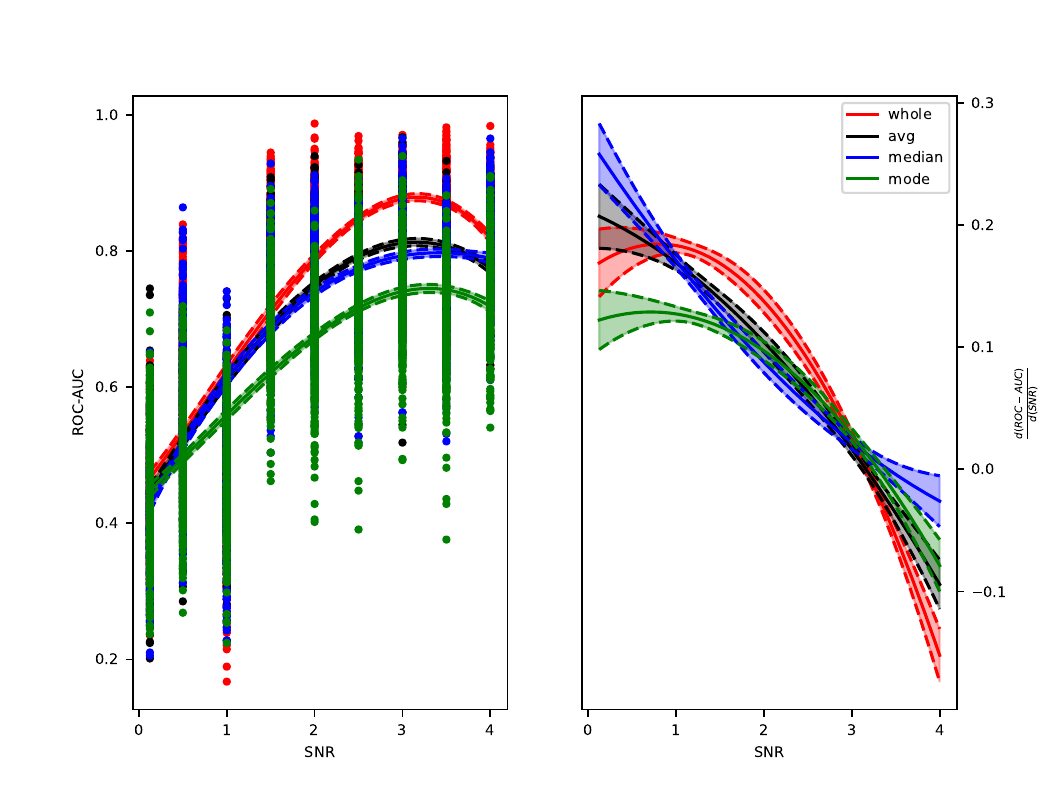}
    \caption{\justifying \footnotesize \textit{Spline fit (left) and local slopes (first derivative- right) of the ROC AUC-SNR data. While at high noise condition ($SNR\leq0.5$), all measures performed similarly at chance level, performance improved with increasing SNRs. }}
    \label{fig:19}
\end{figure}
\justifying \par Figures \ref{fig:16} \& \ref{fig:17} plot the boxplots for accuracies and ROC AUC scores obtained from nested cross validation procedure. As in case of the Rossler attractor example above, we fit smoothened spline regression, and computed the first derivative of these spline curves to quantify the local measures of sensitivity to noise levels. We also estimated the linear slope for the entire distribution (Figures \ref{fig:18} \& \ref{fig:19}) as an overall measure of sensitivity.  

As in case of the Rossler Attractors, we found that under high noise levels ($SNR\leq0.5$) classification using all measures of central tendency and that from the whole RP performed were at chance level, both in terms of accuracy and ROC AUC. At lower SNRs (0.5 to 1.5) a high overlap for the 0.025-0.975 confidence regions of the spline curve is observed for the performance measures obtained from the whole RP, mean and median, indicating an almost identical trend. While at the peak (SNR value of 3), the whole RP performs much better than any other summary statistic, using the overall trend quantified by the linear fits (Figures \ref{fig:18} \& \ref{fig:19}),  the slope for mode was found to be significantly lower than that obtained from other measures suggesting  that the mode is least sensitive to the changes in noise levels, when it comes to performance.

\begin{table} 
\begin{tabular}{ |p{3cm}||p{3cm}|p{3cm}|p{3cm}|  }
 \hline
 \multicolumn{4}{|c|}{Table 4} \\
 \hline
 Estimate & Slope(m) & 2.5\% quantile & 97.5\% quantile \\
 \hline
 Whole RP & 0.09 & 0.086 & 0.092 \\
 \hline 
 mean & 0.072 & 0.068 & 0.075 \\
 \hline 
 median & 0.072 & 0.068 & 0.075 \\
 \hline 
 mode & 0.057 & 0.055 & 0.06 \\
 \hline
\end{tabular}
\caption{\justifying \footnotesize \textit{Global slope of accuracy-SNR data. Among the central tendency measures, mode has the least slope, indicating least sensitivity to the variations in noise level.}}
\label{table-4}
\end{table}

\begin{table} 
\begin{tabular}{ |p{3cm}||p{3cm}|p{3cm}|p{3cm}|  }
 \hline
 \multicolumn{4}{|c|}{Table 5} \\
 \hline
 Estimate & Slope(m) & 2.5\% quantile & 97.5\% quantile \\
 \hline
 Whole RP & 0.103 & 0.099 & 0.107 \\
 \hline 
 mean & 0.090 & 0.086 & 0.094 \\
 \hline 
 median & 0.093 & 0.088 & 0.181 \\
 \hline 
 mode & 0.077 & 0.074 & 0.08 \\
 \hline
\end{tabular}
\caption{\justifying \footnotesize \textit{Global slope of ROC AUC-SNR data. It was found that the mode has the least slope among all central tendency measures estimated, suggesting its low sensitivity to noise. }}
\label{table-5}
\end{table}

\subsection{Test Bed 3: Data From \citeauthor{koul2023interpersonal}(\citeyear{koul2023interpersonal})}
\justifying \par To test our approach on  real-world experimental data, we used openly accessible data from \citeauthor{koul2023interpersonal}(\citeyear{koul2023interpersonal}) who looked at interpersonal synchrony of task free spontaneous movements in dyads sitting across each other under different conditions of visual access (ON or OFF) and proximity (Near vs. Far). Velocity of joints was extracted from video data using OpenPose (\citeauthor{8765346},\citeyear{8765346}; \citeauthor{simon2017hand}, \citeyear{simon2017hand}; \citeauthor{cao2017realtime}, \citeyear{cao2017realtime}; \citeauthor{wei2016cpm}, \citeyear{wei2016cpm}). There were three trials for each of the four visual access-proximity combinations, making it 12 trials per dyad, each of approximately 2 minutes duration. Since the exact length of time series data varied across dyads (range = 73 - 110s, mean = 95.223s, SEM = 0.691s, n = 261), we found it ideal for testing the sliding window MdRQA approach on this dataset and see if it replicates the results reported in the original study. It is important to note that even though MdRQA is well poised to compute recurrence across more than two members in a group, it is well suited to analyze dyad data as well (\citeauthor{wallot2016multidimensional}, \citeyear{wallot2016multidimensional}). \citeauthor{koul2023interpersonal} investigated if interpersonal synchrony of spontaneous body movements in dyads (computed as pairwise correlation between joint movement velocities) varied as a function of visual access and proximity. For the purposes of validating our approach, we chose the velocity data from one particular pair of body part (right foot-right foot) which was reported to show significant correlation between members of a dyad across the vision access ON vs. OFF condition (Fig. 3 and 4 of \citeauthor{koul2023interpersonal}(\citeyear{koul2023interpersonal}), but not in the Near vs. Far condition (Fig. 4 of  \citeauthor{koul2023interpersonal}(\citeyear{koul2023interpersonal}).

\subsubsection{Validation Approach 3: Predicting whether the interpersonal movement synchrony was greater in the a) visual access ON vs. OFF condition and b) near vs. far condition}
\par In the original study, the velocity time series of the right foot-right foot pair across dyads in the dataset was sampled at the rate of 30Hz (corresponding to 120 seconds at max.). Since temporal continuity of data points is important to analyze the changing dynamics of interpersonal movements using our approach, missing data points were interpolated using the piecewise cubic hermite interpolation method (\citeauthor{fritsch1980monotone},\citeyear{fritsch1980monotone}; \citeauthor{kahaner1989numerical},\citeyear{kahaner1989numerical}). Data was then smoothened using a moving average procedure (window length of 30 samples = 1s) as reported in the original study and downsampled to 5Hz. Euclidian magnitude of velocity was then computed from x and y coordinate values in the data. We computed the MdRQA variables using the sliding window approach for both the original time series (see Supplementary Information for more details) as well as its randomly permuted version. The latter was used as a control condition to compare with the classifier performance of the original data, since perturbing the temporal structure of the time series would be expected to yield performance at chance. Based on the results of the validation approaches 1 and 2 reported in the earlier sections of this study, we used the mode of the MdRQA variable distribution across the sliding windows, after z-transformation (since we were interested in the relative (and not the absolute) differences in the MdRQA variable values across dyads) for the nested cross validation procedure described above. Please see Supplementary Information for results pertaining to classifier performance derived using other measures of central tendency as well as the whole RP (Supplementary figures \ref{fig:S4}, \ref{fig:S5}, \ref{fig:S6}, \ref{fig:S7}, \ref{fig:S8}, \ref{fig:S9}).

\subsubsection{Results of Validation Approach 3}
\begin{figure} 
    \centering
    \includegraphics[width=\textwidth]{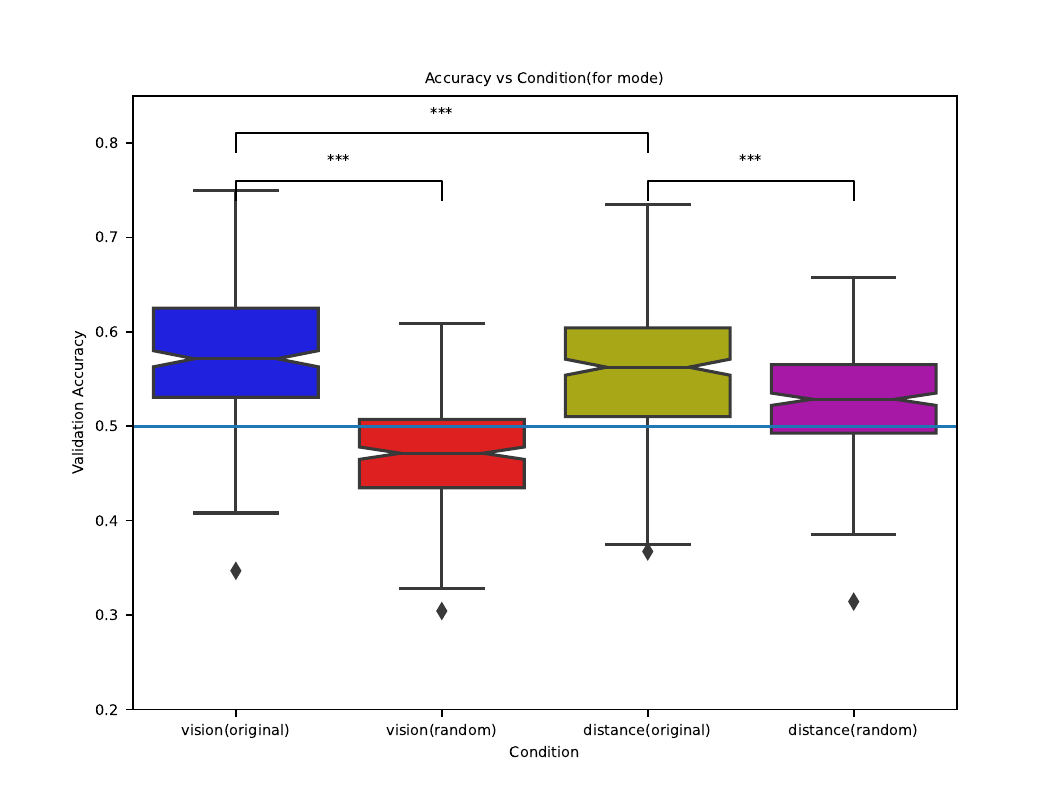}
    \caption{\justifying \footnotesize \textit{Box plots of classifier accuracy estimates on validation sets from nested cross validation using mode of the MdRQA variable distributions across the sliding windows, predicting whether visual access between participants in a dyad was ON vs. OFF using the original (a) and the randomized (b) time series and whether the two participants in a dyad were seated near vs. far using the original (c) and the randomized (d) time series. Horizontal line at y = 0.5 indicates classifier performance at chance. ns = non significant difference, * = p<$\alpha$, ** = p<$\alpha/5$, *** = p<$\alpha/50$}}
    \label{fig:20}
\end{figure}
\begin{figure} 
    \centering
    \includegraphics[width=\textwidth]{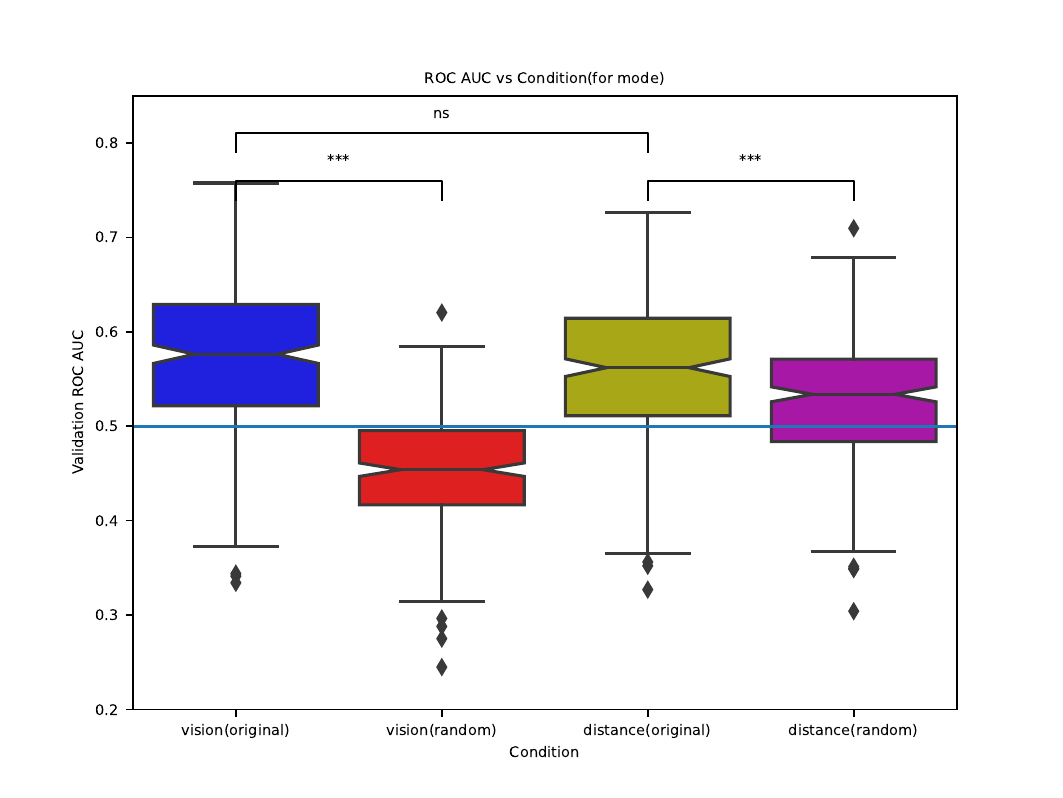}
    \caption{\justifying \footnotesize \textit{Box plots of classifier ROC AUC estimates on validation sets from nested cross validation using mode of the MdRQA variable distributions across the sliding windows, predicting whether visual access between participants in a dyad was ON vs. OFF using the original (a) and  randomized (b) time series, and whether individual participants in a dyad were seated near vs. far using the original (c) and the randomized (d) time series. Horizontal line at y = 0.5 indicates classifier performance at chance. ns = non significant difference, * = p<$\alpha$, ** = p<$\alpha/5$, *** = p<$\alpha/50$}}
    \label{fig:21}
\end{figure}
Figures 20 and 21 plot the boxplots for classifier accuracy and ROC AUC values obtained from nested cross validation procedure. As expected from the results reported in the original study, we found that the classifier accuracy as well as the ROC AUC values for the visual access ON vs. OFF classification, using the mode of the MdRQA variable distribution across sliding windows, was significantly higher in the original (median accuracy = 0.572, median ROC AUC = 0.576) vs. randomized (median accuracy = 0.472, median ROC AUC = 0.454) time series (classifier accuracy: Wilcoxon signed rank test (one sided), Bonferroni corrected $\alpha$=0.017, $w$ = 4.36e+4, p = 9.786e-45, $z$ = 13.734, effect size= 0.807; ROC AUC: Wilcoxon signed rank test (one sided), Bonferroni corrected $\alpha$=0.017, $w$ = 4.33e+4, p = 1.307e-43, $z$ = 13.798, effect size= 0.797). However, in contrast to that reported in the original study, we also found the classifier accuracy as well as ROC AUC values for predicting whether the participants in a dyad were seated near vs. far, using the velocity data of the right foot-right foot pair, to be significantly higher in the original (median accuracy = 0.562, median ROC AUC = 0.562) vs. the randomized (median accuracy = 0.528, median ROC AUC = 0.533) time series (classifier accuracy: Wilcoxon signed rank test (one sided), Bonferroni corrected $\alpha$=0.017, $w$ = 3.092e+4, p = 1.419e-8, $z$ = 5.551, effect size= 0.320; and ROC AUC: Wilcoxon signed rank test (one sided), Bonferroni corrected $\alpha$=0.017, $w$ = 3.126e+4, p = 3.834e-9, $z$ = 5.775, effect size= 0.333). While the significantly higher median values of classifier accuracy (and not ROC AUC values) for predicting visual access conditions when compared to proximity conditions is consistent with the overall trends observed in the original study (classifier accuracy: Wilcoxon signed rank test (one sided), Bonferroni corrected $\alpha$=0.017, $w$ = 2.457e+4, p = 1.506e-4, $z$ = 1.327, effect size= 0.0766; ROC AUC: Wilcoxon signed rank test (one sided), Bonferroni corrected $\alpha$=0.017, Wilcoxon signed rank test (one sided), $w$ = 2.5e+4, p = 5.341e-2, $z$ = 1.613, effect size= 0.0931),  given the discrepancy in the proximity results between ours and the original study, we decided to test the proximity classifier performance separately in the visual access ON vs. OFF conditions. While in the visual access ON condition, the Near vs. Far classifier using the mode of the MdRQA variable distribution across the sliding windows was still performing significantly above chance (median accuracy = 0.63, Wilcoxon signed rank test (one sided), $w$ = 4.13e+4.0, p = 5.226e-36, $z$ = 12.446, effect size= 0.718)  (Supplementary figures. \ref{fig:S10} \& \ref{fig:S11}), it was not so in the visual access OFF condition (median accuracy = 0.476, Wilcoxon signed rank test (one sided), $w$ = 9.408e+3.0, p = 1.0, $z$ = -8.756, effect size= -0.505), suggesting that the performance in predicting the proximity predictions was indeed dependent on visual access as indicated by the results of the original study.

\section{Discussion}
\justifying \par In recent years, RQA and its multidimensional version (MdRQA) have emerged as a popular tool for assessing interpersonal behavioral or physiological synchrony across individuals in a variety of contexts, owing to the general robustness of the method to non-linearity in data. While experimental data in such studies are typically collected for a fixed, pre-determined duration, a systematic attempt validating its use in comparing synchrony between samples of non-uniform composition and unequal time series data has been lacking. In this study, we proposed a  sliding window MdRQA approach to address this and tested the ability of our approach in discriminating between two dynamic states of a complex system, when the time series data from samples varied in composition and/ or duration. We used two well-known dynamical systems (the Rossler Attractor and the Kuramoto system of coupled oscillators) and openly accessible data from a previously published study as the testbed for validating our approach. 

In the case of Rossler Attractors, we found that under conditions of variable RP durations (150-450 equivalent to 600-1800s) and low signal to noise ratios (at $SNR \geq1$), all three measures of central tendency (mean, mode and median) of the MdRQA variable distributions across sliding windows of a RP, could discriminate between the chaotic vs. periodic state of the attractor. In particular, mode of the distributions performed at or above chance but always better than the whole RP estimates. Moreover, while the classifier performance measures increased with SNR values, the mode of the MdRQA variable distributions (when compared to the mean or median) varied the least for the same amount of change in SNR. This suggests that the mode of the MdRQA variable distributions could be a conservative but robust measure of classifier performance, when the noise levels are unknown, as is usually the case in most empirical studies. 

\justifying \par Since the number of dimensions is fixed in the Rossler system,  we used the Kuramoto system of coupled oscillators to examine the efficacy of our approach against a varying number of observable variables (components or dimensions). Under varying levels of noise, RP durations (150-450) and number of oscillators (3-6), we again found central tendencies of the sliding window distributions to be able to classify whether the coupled oscillators would be in sync or not, based on the coupling strength. In particular, all measures of central tendency  had equivalent discriminative accuracies in the high noise condition ($SNR \leq0.5$) though mode was again found to be least sensitive to levels of noise in the signal. It is unclear why mode had the least variation against noise levels in both the systems we examined via simulation in this study. It  may even be possible that for another system, the level of accuracy at which the mode performs (even if better than other measures), is too low to be practically desirable.  However, it still may be a measure of choice when opting for a conservative criterion. In other words, while we caution against concluding that mode is universally the best choice for all systems, it still may be a safer option when one is uncertain about the noise levels in the system.

\justifying \par For the purpose of validating our approach, the Kuramoto system of coupled oscillators was chosen in its most simplified form (the mean field model) even though real-world data could be expected to exhibit a much richer dynamics, say, from systems with multiple interacting, inter-dependent components. In fact, it would be possible to theoretically incorporate varying levels of noise to the extent of coupling between any two oscillators to further simulate this gradient of inter-dependency expected in real world systems and then test the usefulness of the approach and generalizability of the inferences drawn. However, we considered real world experimental data to be a stricter testbed for the validity of our approach and hence, proceeded to analyze the openly accessible data from a recently published study looking at task-free interpersonal movement synchrony in dyads (\citeauthor{koul2023interpersonal}(\citeyear{koul2023interpersonal}). In particular, we analyzed the movement velocity of the the right foot-right foot pair of members in a dyad across varying conditions of visual access (ON vs. OFF) and proximity (Near vs. Far). While the original study, using pairwise correlations as a measure of synchrony, had reported a significant main effect of visual access, but not of proximity, in this pair of body parts, we found the mode of the MdRQA variable distributions across sliding windows to classify visual access as well as proximity conditions above chance and significantly greater than predicted by the randomized versions of the respective time series (Fig. \ref{fig:20} \& \ref{fig:21}). The discrepancy in the proximity results was clarified when the proximity classification was done for the visual access ON and OFF datasets separately and classifier performance for Near vs. Far was found to be significant only in the former case. These results are consistent with the overall trends reported in the original study (specifically, seen for the left foot-left knee pair of body parts) suggesting that MdRQA could perhaps be a more sensitive read-out of synchrony than linear, correlation based methods (\citeauthor{young2015we}, \citeyear{young2015we}). In sum, using this dataset as the third testbed for our approach, allowed us to successfully validate and extend the applicability of our proposed method to complex, real world contexts.

\justifying \par In the current study, the sliding window MdRQA method was tested in simple cases where the classification required was a binary one. Hence, the applicability  of our approach to scenarios where there could be more than two categories of system dynamics to be classified remains to be tested. Lastly, the noise  that was introduced in both  models under study was Gaussian white noise which is  only one type of noise that could be expected in a system. In addition, the SNR range  was arbitrarily chosen to elicit a range of classifier performance across noise levels, and mostly until the performance is getting saturated. In the absence of a  functional form to the relationship between the SNR and performance measure,  the inferences drawn may be limited to the range of noise levels we tested. However, despite these limitations, we believe that this work is the first systematic attempt to allow drawing some useful insights and paving the way for future work that examines interpersonal synchrony in more naturalistic, ecologically valid contexts.  

\section{Conclusion}
In this study, we proposed a sliding window MdRQA approach to compute and compare synchrony across multi-dimensional samples of unequal duration and composition. We tested and validated this approach first via simulations of two widely studied non-linear dynamical systems and then using real world experimental data examining interpersonal movement synchrony. We found that measures of central tendency of the MdRQA variable distributions across the sliding windows, could classify the dynamic states of a sample of systems, with a cross validation accuracy significantly greater than chance. In addition, compared to other measures of central tendency, mode was found to be the most robust to changes in noise levels in the simulated data. 

\section{Acknowledgments}
We thank the Remote Computing Facility at the Department of Cognitive Science, IIT Kanpur for their support.

\section{Code Availability}
\url{https://github.com/SwaragThaikkandi/Sliding_MdRQA}




\printbibliography

@article{guastello2006electrodermal,
  title={Electrodermal arousal between participants in a conversation: nonlinear dynamics and linkage effects.},
  author={Guastello, Stephen J and Pincus, David and Gunderson, Patrick R},
  journal={Nonlinear dynamics, psychology, and life sciences},
  year={2006},
  publisher={Society for Chaos Theory in Psychology \& Life Sciences}
}

@article{richardson2007art,
  title={The art of conversation is coordination},
  author={Richardson, Daniel C and Dale, Rick and Kirkham, Natasha Z},
  journal={Psychological science},
  volume={18},
  number={5},
  pages={407--413},
  year={2007},
  publisher={SAGE Publications Sage CA: Los Angeles, CA}
}

@article{shockley2003mutual,
  title={Mutual interpersonal postural constraints are involved in cooperative conversation.},
  author={Shockley, Kevin and Santana, Marie-Vee and Fowler, Carol A},
  journal={Journal of Experimental Psychology: Human Perception and Performance},
  volume={29},
  number={2},
  pages={326},
  year={2003},
  publisher={American Psychological Association}
}

@article{konvalinka2011synchronized,
  title={Synchronized arousal between performers and related spectators in a fire-walking ritual},
  author={Konvalinka, Ivana and Xygalatas, Dimitris and Bulbulia, Joseph and Schj{\o}dt, Uffe and Jegind{\o}, Else-Marie and Wallot, Sebastian and Van Orden, Guy and Roepstorff, Andreas},
  journal={Proceedings of the National Academy of Sciences},
  volume={108},
  number={20},
  pages={8514--8519},
  year={2011},
  publisher={National Acad Sciences}
}

@article{yun2012interpersonal,
  title={Interpersonal body and neural synchronization as a marker of implicit social interaction},
  author={Yun, Kyongsik and Watanabe, Katsumi and Shimojo, Shinsuke},
  journal={Scientific reports},
  volume={2},
  number={1},
  pages={959},
  year={2012},
  publisher={Nature Publishing Group UK London}
}

@article{hasson2012brain,
  title={Brain-to-brain coupling: a mechanism for creating and sharing a social world},
  author={Hasson, Uri and Ghazanfar, Asif A and Galantucci, Bruno and Garrod, Simon and Keysers, Christian},
  journal={Trends in cognitive sciences},
  volume={16},
  number={2},
  pages={114--121},
  year={2012},
  publisher={Elsevier}
}

@article{spiegelhalder2014interindividual,
  title={Interindividual synchronization of brain activity during live verbal communication},
  author={Spiegelhalder, Kai and Ohlendorf, Sabine and Regen, Wolfram and Feige, Bernd and van Elst, Ludger Tebartz and Weiller, Cornelius and Hennig, J{\"u}rgen and Berger, Mathias and T{\"u}scher, Oliver},
  journal={Behavioural brain research},
  volume={258},
  pages={75--79},
  year={2014},
  publisher={Elsevier}
}

@article{golland2015mere,
  title={The mere co-presence: Synchronization of autonomic signals and emotional responses across co-present individuals not engaged in direct interaction},
  author={Golland, Yulia and Arzouan, Yossi and Levit-Binnun, Nava},
  journal={PloS one},
  volume={10},
  number={5},
  pages={e0125804},
  year={2015},
  publisher={Public Library of Science San Francisco, CA USA}
}

@article{mitkidis2015building,
  title={Building trust: Heart rate synchrony and arousal during joint action increased by public goods game},
  author={Mitkidis, Panagiotis and McGraw, John J and Roepstorff, Andreas and Wallot, Sebastian},
  journal={Physiology \& behavior},
  volume={149},
  pages={101--106},
  year={2015},
  publisher={Elsevier}
}

@article{wallot2016multidimensional,
  title={Multidimensional Recurrence Quantification Analysis (MdRQA) for the analysis of multidimensional time-series: A software implementation in MATLAB and its application to group-level data in joint action},
  author={Wallot, Sebastian and Roepstorff, Andreas and M{\o}nster, Dan},
  journal={Frontiers in psychology},
  pages={1835},
  year={2016},
  publisher={Frontiers}
}

@article{bevilacqua2019brain,
  title={Brain-to-brain synchrony and learning outcomes vary by student--teacher dynamics: Evidence from a real-world classroom electroencephalography study},
  author={Bevilacqua, Dana and Davidesco, Ido and Wan, Lu and Chaloner, Kim and Rowland, Jess and Ding, Mingzhou and Poeppel, David and Dikker, Suzanne},
  journal={Journal of cognitive neuroscience},
  volume={31},
  number={3},
  pages={401--411},
  year={2019},
  publisher={MIT Press One Rogers Street, Cambridge, MA 02142-1209, USA journals-info~…}
}

@article{pan2020instructor,
  title={Instructor-learner brain coupling discriminates between instructional approaches and predicts learning},
  author={Pan, Yafeng and Dikker, Suzanne and Goldstein, Pavel and Zhu, Yi and Yang, Cuirong and Hu, Yi},
  journal={NeuroImage},
  volume={211},
  pages={116657},
  year={2020},
  publisher={Elsevier}
}

@article{takens1981dynamical,
  title={Dynamical systems and turbulence},
  author={Takens, Floris},
  journal={Warwick, 1980},
  pages={366--381},
  year={1981},
  publisher={Springer-Verlag}
}

@article{kennel1992determining,
  title={Determining embedding dimension for phase-space reconstruction using a geometrical construction},
  author={Kennel, Matthew B and Brown, Reggie and Abarbanel, Henry DI},
  journal={Physical review A},
  volume={45},
  number={6},
  pages={3403},
  year={1992},
  publisher={APS}
}

@article{hegger1999improved,
  title={Improved false nearest neighbor method to detect determinism in time series data},
  author={Hegger, Rainer and Kantz, Holger},
  journal={Physical Review E},
  volume={60},
  number={4},
  pages={4970},
  year={1999},
  publisher={APS}
}

@article{marwan2013recurrence,
  title={Recurrence plots 25 years later—Gaining confidence in dynamical transitions},
  author={Marwan, Norbert and Schinkel, Stefan and Kurths, J{\"u}rgen},
  journal={Europhysics Letters},
  volume={101},
  number={2},
  pages={20007},
  year={2013},
  publisher={IOP Publishing}
}

@book{huffaker2017nonlinear,
  title={Nonlinear time series analysis with R},
  author={Huffaker, Ray and Huffaker, Ray G and Bittelli, Marco and Rosa, Rodolfo},
  year={2017},
  publisher={Oxford University Press}
}

@article{kaufman2012leakage,
  title={Leakage in data mining: Formulation, detection, and avoidance},
  author={Kaufman, Shachar and Rosset, Saharon and Perlich, Claudia and Stitelman, Ori},
  journal={ACM Transactions on Knowledge Discovery from Data (TKDD)},
  volume={6},
  number={4},
  pages={1--21},
  year={2012},
  publisher={ACM New York, NY, USA}
}

@article{rossler1976equation,
  title={An equation for continuous chaos},
  author={R{\"o}ssler, Otto E},
  journal={Physics Letters A},
  volume={57},
  number={5},
  pages={397--398},
  year={1976},
  publisher={Elsevier}
}

@article{ermentrout1991adaptive,
  title={An adaptive model for synchrony in the firefly Pteroptyx malaccae},
  author={Ermentrout, Bard},
  journal={Journal of Mathematical Biology},
  volume={29},
  number={6},
  pages={571--585},
  year={1991},
  publisher={Springer}
}

@article{thorson2021oxytocin,
  title={Oxytocin increases physiological linkage during group therapy for methamphetamine use disorder: a randomized clinical trial},
  author={Thorson, Katherine R and McKernan, Scott M and West, Tessa V and Woolley, Joshua D and Mendes, Wendy Berry and Stauffer, Christopher S},
  journal={Scientific reports},
  volume={11},
  number={1},
  pages={21004},
  year={2021},
  publisher={Nature Publishing Group UK London}
}

@article{gordon2021group,
  title={Group-level physiological synchrony and individual-level anxiety predict positive affective behaviors during a group decision-making task},
  author={Gordon, Ilanit and Wallot, Sebastian and Berson, Yair},
  journal={Psychophysiology},
  volume={58},
  number={9},
  pages={e13857},
  year={2021},
  publisher={Wiley Online Library}
}

@article{tomashin2022interpersonal,
  title={Interpersonal physiological synchrony predicts group cohesion},
  author={Tomashin, Alon and Gordon, Ilanit and Wallot, Sebastian},
  journal={Frontiers in human neuroscience},
  volume={16},
  pages={903407},
  year={2022},
  publisher={Frontiers}
}

@article{marwan2007recurrence,
  title={Recurrence plots for the analysis of complex systems},
  author={Marwan, Norbert and Romano, M Carmen and Thiel, Marco and Kurths, J{\"u}rgen},
  journal={Physics reports},
  volume={438},
  number={5-6},
  pages={237--329},
  year={2007},
  publisher={Elsevier}
}

@article{webber1994dynamical,
  title={Dynamical assessment of physiological systems and states using recurrence plot strategies},
  author={Webber Jr, Charles L and Zbilut, Joseph P},
  journal={Journal of applied physiology},
  volume={76},
  number={2},
  pages={965--973},
  year={1994}
}

@article{marwan2002recurrence,
  title={Recurrence-plot-based measures of complexity and their application to heart-rate-variability data},
  author={Marwan, Norbert and Wessel, Niels and Meyerfeldt, Udo and Schirdewan, Alexander and Kurths, J{\"u}rgen},
  journal={Physical review E},
  volume={66},
  number={2},
  pages={026702},
  year={2002},
  publisher={APS}
}

@article{schippers2010mapping,
  title={Mapping the information flow from one brain to another during gestural communication},
  author={Schippers, Marleen B and Roebroeck, Alard and Renken, Remco and Nanetti, Luca and Keysers, Christian},
  journal={Proceedings of the National Academy of Sciences},
  volume={107},
  number={20},
  pages={9388--9393},
  year={2010},
  publisher={National Acad Sciences}
}

@article{stuldreher2023robustness,
  title={Robustness of Physiological Synchrony in Wearable Electrodermal Activity and Heart Rate as a Measure of Attentional Engagement to Movie Clips},
  author={Stuldreher, Ivo V and van Erp, Jan BF and Brouwer, Anne-Marie},
  journal={Sensors},
  volume={23},
  number={6},
  pages={3006},
  year={2023},
  publisher={MDPI}
}

@article{webber2005recurrence,
  title={Recurrence quantification analysis of nonlinear dynamical systems},
  author={Webber Jr, Charles L and Zbilut, Joseph P},
  journal={Tutorials in contemporary nonlinear methods for the behavioral sciences},
  volume={94},
  number={2005},
  pages={26--94},
  year={2005}
}

@article{webber1995influence,
  title={Influence of isometric loading on biceps EMG dynamics as assessed by linear and nonlinear tools},
  author={Webber Jr, Charles L and Schmidt, MA and Walsh, John M},
  journal={Journal of Applied Physiology},
  volume={78},
  number={3},
  pages={814--822},
  year={1995}
}

@article{dikker2017brain,
  title={Brain-to-brain synchrony tracks real-world dynamic group interactions in the classroom},
  author={Dikker, Suzanne and Wan, Lu and Davidesco, Ido and Kaggen, Lisa and Oostrik, Matthias and McClintock, James and Rowland, Jess and Michalareas, Georgios and Van Bavel, Jay J and Ding, Mingzhou and others},
  journal={Current biology},
  volume={27},
  number={9},
  pages={1375--1380},
  year={2017},
  publisher={Elsevier}
}

@article{wallot2018calculation,
  title={Calculation of average mutual information (AMI) and false-nearest neighbors (FNN) for the estimation of embedding parameters of multidimensional time series in matlab},
  author={Wallot, Sebastian and M{\o}nster, Dan},
  journal={Frontiers in psychology},
  volume={9},
  pages={1679},
  year={2018},
  publisher={Frontiers Media SA}
}

@article{verstynen2023overfitting,
  title={Overfitting to ‘predict’suicidal ideation},
  author={Verstynen, Timothy and Kording, Konrad Paul},
  journal={Nature Human Behaviour},
  volume={7},
  number={5},
  pages={680--681},
  year={2023},
  publisher={Nature Publishing Group UK London}
}

@article{acebron2005kuramoto,
  title={The Kuramoto model: A simple paradigm for synchronization phenomena},
  author={Acebr{\'o}n, Juan A and Bonilla, Luis L and Vicente, Conrad J P{\'e}rez and Ritort, F{\'e}lix and Spigler, Renato},
  journal={Reviews of modern physics},
  volume={77},
  number={1},
  pages={137},
  year={2005},
  publisher={APS}
}

@article{kuramoto1975international,
  title={International symposium on mathematical problems in theoretical physics},
  author={Kuramoto, Yoshiki},
  journal={Lecture notes in Physics},
  volume={30},
  pages={420},
  year={1975}
}

@article{chopra2009exponential,
  title={On exponential synchronization of Kuramoto oscillators},
  author={Chopra, Nikhil and Spong, Mark W},
  journal={IEEE transactions on Automatic Control},
  volume={54},
  number={2},
  pages={353--357},
  year={2009},
  publisher={IEEE}
}

@article{strogatz2000kuramoto,
  title={From Kuramoto to Crawford: exploring the onset of synchronization in populations of coupled oscillators},
  author={Strogatz, Steven H},
  journal={Physica D: Nonlinear Phenomena},
  volume={143},
  number={1-4},
  pages={1--20},
  year={2000},
  publisher={Elsevier}
}

@article{biccari2020stochastic,
  title={A stochastic approach to the synchronization of coupled oscillators},
  author={Biccari, Umberto and Zuazua, Enrique},
  journal={Frontiers in Energy Research},
  volume={8},
  pages={115},
  year={2020},
  publisher={Frontiers Media SA}
}

@book{kantz2004nonlinear,
  title={Nonlinear time series analysis},
  author={Kantz, Holger and Schreiber, Thomas},
  volume={7},
  year={2004},
  publisher={Cambridge university press}
}

@article{eckmann1995recurrence,
  title={Recurrence plots of dynamical systems},
  author={Eckmann, Jean-Pierre and Kamphorst, S Oliffson and Ruelle, David and others},
  journal={World Scientific Series on Nonlinear Science Series A},
  volume={16},
  pages={441--446},
  year={1995},
  publisher={World Scientific Publishing}
}

@book{strogatz2018nonlinear,
  title={Nonlinear dynamics and chaos with student solutions manual: With applications to physics, biology, chemistry, and engineering},
  author={Strogatz, Steven H},
  year={2018},
  publisher={CRC press}
}

@article{koul2023interpersonal,
  title={Interpersonal synchronization of spontaneously generated body movements},
  author={Koul, Atesh and Ahmar, Davide and Iannetti, Gian Domenico and Novembre, Giacomo},
  journal={Iscience},
  volume={26},
  number={3},
  year={2023},
  publisher={Elsevier}
}

@article{schreiber1996improved,
  title={Improved surrogate data for nonlinearity tests},
  author={Schreiber, Thomas and Schmitz, Andreas},
  journal={Physical review letters},
  volume={77},
  number={4},
  pages={635},
  year={1996},
  publisher={APS}
}

@article{kugiumtzis1999test,
  title={Test your surrogate data before you test for nonlinearity},
  author={Kugiumtzis, D},
  journal={Physical Review E},
  volume={60},
  number={3},
  pages={2808},
  year={1999},
  publisher={APS}
}

@article{schreiber1997discrimination,
  title={Discrimination power of measures for nonlinearity in a time series},
  author={Schreiber, Thomas and Schmitz, Andreas},
  journal={Physical Review E},
  volume={55},
  number={5},
  pages={5443},
  year={1997},
  publisher={APS}
}

@article{fritsch1980monotone,
  title={Monotone piecewise cubic interpolation},
  author={Fritsch, Frederick N and Carlson, Ralph E},
  journal={SIAM Journal on Numerical Analysis},
  volume={17},
  number={2},
  pages={238--246},
  year={1980},
  publisher={SIAM}
}

@book{kahaner1989numerical,
  title={Numerical methods and software},
  author={Kahaner, David and Moler, Cleve and Nash, Stephen},
  year={1989},
  publisher={Prentice-Hall, Inc.}
}

@article{8765346,
  author = {Z. {Cao} and G. {Hidalgo Martinez} and T. {Simon} and S. {Wei} and Y. A. {Sheikh}},
  journal = {IEEE Transactions on Pattern Analysis and Machine Intelligence},
  title = {OpenPose: Realtime Multi-Person 2D Pose Estimation using Part Affinity Fields},
  year = {2019}
}

@inproceedings{simon2017hand,
  author = {Tomas Simon and Hanbyul Joo and Iain Matthews and Yaser Sheikh},
  booktitle = {CVPR},
  title = {Hand Keypoint Detection in Single Images using Multiview Bootstrapping},
  year = {2017}
}

@inproceedings{cao2017realtime,
  author = {Zhe Cao and Tomas Simon and Shih-En Wei and Yaser Sheikh},
  booktitle = {CVPR},
  title = {Realtime Multi-Person 2D Pose Estimation using Part Affinity Fields},
  year = {2017}
}

@inproceedings{wei2016cpm,
  author = {Shih-En Wei and Varun Ramakrishna and Takeo Kanade and Yaser Sheikh},
  booktitle = {CVPR},
  title = {Convolutional pose machines},
  year = {2016}
}

@article{young2015we,
  title={We should be using nonlinear indices when relating heart-rate dynamics to cognition and mood},
  author={Young, Hayley and Benton, David},
  journal={Scientific reports},
  volume={5},
  number={1},
  pages={16619},
  year={2015},
  publisher={Nature Publishing Group UK London}
}

\appendix*

\justifying 
\newpage
\part*{Supplementary Information}
\addcontentsline{toc}{part}{Supplementary Information}
\thispagestyle{empty} 

\title{Supplementary Information}
\section{Perturbation Analysis For Rossler Attractor}
\renewcommand{\theequation}{SA.\arabic{equation}}
\justifying \par In addition to numerical estimation, a perturbation analysis was carried out to confirm the bifurcation of a Rossler Attractor at $a = 0.2$. A perturbation is a small change in the initial conditions which if grows exponentially over time, the system is said to be chaotic. 

The time series data was simulated using 30 different initial states and perturbed by the same small amount (0.001 unit) in any of the three dimensions randomly. The perturbation was sampled as the surface of a sphere (with radius, $r (=0.001)$) whose center was taken to be the original initial point. The coordinates of a point on the surface of a sphere would have the form:
\begin{align}
\begin{split}
x &= x_{0} +  r \cdot \sin(\theta) \cdot \cos(\phi) \\
y &= y_{0} +  r \cdot \sin(\theta) \cdot \sin(\phi) \\
z &= z_{0} +  r \cdot \cos(\theta)
\end{split}
\end{align}
where:
\begin{itemize}
    \item $r$ is the radial distance from the origin to a point on the sphere,
    \item $\theta$ is the polar angle (angle between the positive $z$-axis and the line connecting the origin to the point),
    \item $\phi$ is the azimuthal angle measured from the positive $x$-axis in the $xy$-plane.
    \item $x_{0}$, $y_{0}$ \& $z_{0}$ are the initial state coordinates of the actual time series(which wasn't parturbated)
\end{itemize}

The values of $\theta$ and $\phi$ were randomly sampled from a uniform distribution between 0 and $2\pi$.

Following a small change in the initial state (known as perturbation), the Euclidean distance between states for the same initial time point was computed across the length of the time series. A sensitive dependence on initial condition is a characteristic of chaotic systems. A exponential growth would result in a positive Lyapnov exponent (\citeauthor{strogatz2018nonlinear}, \citeyear{strogatz2018nonlinear}, section 10.5), on which when a logarithm is applied, a linear region is obtained. Fig. \ref{fig:3} for $a>=0.2$  showing that for values in that range, the system is indeed showing chaotic behaviour. 
\begin{suppfigure}
    \centering
    \includegraphics[width=\linewidth]{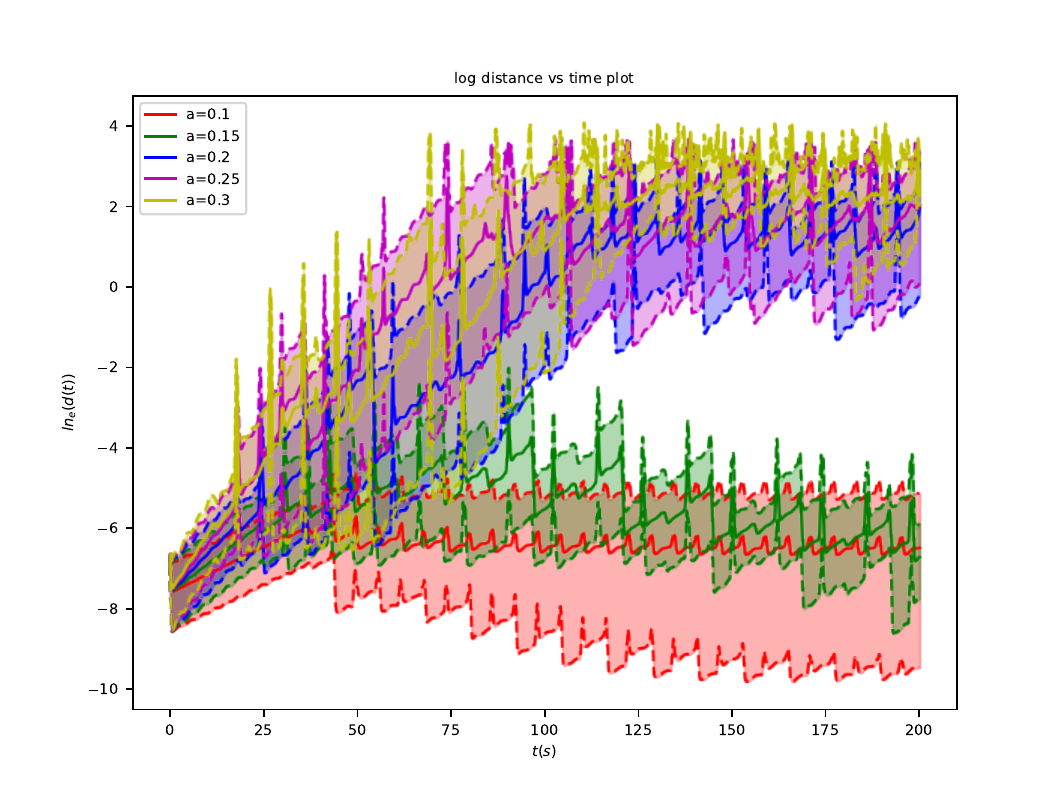}
    \caption{\justifying \footnotesize \textit{Log of distance between points on actual time series and perturbed time series at each time point for the phase space trajectories plotted in a, b, c \& d above. This is done by defining a sphere having radius the amount of perturbation around the initial point of the time series and putting points on the surface of the sphere as initial points for the perturbed time series. For chaotic systems, this distance, that we introduced initially would grow exponentially over time, resulting in a linear increase in the log graph (positive Lyapnov exponent). But, for periodic systems, this distance may reduce or stay the same. This graph verifies that the system was showing chaotic behaviour for $a\geq0.2$}}
    \label{fig:3}
\end{suppfigure}

\section{Time Delayed Embedding}\label{section:tde}
\renewcommand{\theequation}{SB.\arabic{equation}}

\citeauthor{takens1981dynamical}(\citeyear{takens1981dynamical}) had proved that for a complex system, if we have access to only one variable, the whole dynamics of the system could be reconstructed by plotting the same variable against it for a certain number of times with a certain time delay.

Let the observable time series be x, given by: 

\begin{equation}
x= ( x_{1}, x_{2}, x_{3}, ... , x_{n})
\end{equation}
Now using time delayed embedding, if we assume that the embedding dimension is m, we can construct the multidimensional vector for a point as follows:
\begin{equation}
\vec{x}_{1} = ( x_{1} , x_{1+\tau}, x_{1+2\tau}, x_{1+3\tau}, ... , x_{1+(m-1)\tau}) 
\end{equation}
now we can repeat the process by starting the next vector from $x_{2}$ and we can continue until $x_{n-(m-1)\tau}$ as follows: 
\begin{equation}
\bar{X}=
\begin{pmatrix}
     \text{ $\vec{x}_{1}$}\\
     \text{ $\vec{x}_{2}$}\\
     \text{ $\vec{x}_{3}$}\\
     \text{ $.$}\\
     \text{ $.$}\\
     \text{ $.$}\\
     \text{$\vec{x}_{n-(m-1)\tau}$}
\end{pmatrix}
= 
\begin{pmatrix}
\text{$x_{1} , x_{1+\tau}, x_{1+2\tau}, x_{1+3\tau}, ... , x_{1+(m-1)\tau}$} \\
\text{$x_{2} , x_{2+\tau}, x_{2+2\tau}, x_{2+3\tau}, ... , x_{2+(m-1)\tau}$} \\
\text{$x_{3} , x_{3+\tau}, x_{3+2\tau}, x_{3+3\tau}, ... , x_{3+(m-1)\tau}$} \\
\text{ $.$}\\
\text{ $.$}\\
\text{ $.$}\\
\text{$x_{n-(m-1)\tau} , x_{n-(m-2)\tau}, x_{n-(m-3)\tau}, x_{n-(m-4)\tau}, ... , x_{n}$} \\
\end{pmatrix}
\end{equation}

In our case the time series itself was multidimensional, which could be written as:
\begin{equation}
\vec{x}=
\begin{pmatrix}
\text{$x_{1,1}, x_{1,2}, x_{1,3}, ..., x_{1,N}$} \\
\text{$x_{2,1}, x_{2,2}, x_{2,3}, ..., x_{2,N}$} \\
\text{$x_{3,1}, x_{3,2}, x_{3,3}, ..., x_{3,N}$} \\
\text{ $.$}\\
\text{ $.$}\\
\text{ $.$}\\
\text{$x_{n,1}, x_{n,2}, x_{n,3}, ..., x_{n,N}$} \\
\end{pmatrix}
\end{equation}
Where N is the number of members in the group and n is the length of the time series.
Then the multidimensional vector for a point would be: 
\begin{equation}
\vec{x}_{1}= 
\begin{pmatrix}
\text{$\left(x_{1,1}, x_{1,2}, ..., x_{1,N}\right),  \left( x_{1+\tau,1}, x_{1+\tau,2}, ..., x_{1+\tau,N}\right), ... \left(x_{1+(m-1)\tau,1}, x_{1+(m-1)\tau,2}, ..., x_{1+(m-1)\tau,N}\right)$}
\end{pmatrix}
\end{equation}
Then the reconstruction would look like
\begin{equation}
\vec{X}_{1}= 
\begin{pmatrix}
\text{$\left(x_{1,1}, x_{1,2}, ..., x_{1,N}\right),  \left( x_{1+\tau,1}, x_{1+\tau,2}, ..., x_{1+\tau,N}\right), ... \left(x_{1+(m-1)\tau,1}, x_{1+(m-1)\tau,2}, ..., x_{1+(m-1)\tau,N}\right)$}\\
\text{$\left(x_{2,1}, x_{2,2}, ..., x_{2,N}\right),  \left( x_{2+\tau,1}, x_{2+\tau,2}, ..., x_{2+\tau,N}\right), ... \left(x_{2+(m-1)\tau,1}, x_{2+(m-1)\tau,2}, ..., x_{2+(m-1)\tau,N}\right)$}\\
\text{$\left(x_{3,1}, x_{3,2}, ..., x_{3,N}\right),  \left( x_{3+\tau,1}, x_{3+\tau,2}, ..., x_{3+\tau,N}\right), ... \left(x_{3+(m-1)\tau,1}, x_{3+(m-1)\tau,2}, ..., x_{3+(m-1)\tau,N}\right)$}\\
\text{ $.$}\\
\text{ $.$}\\
\text{ $.$}\\
\text{$\left(x_{n-(m-1)\tau,1}, x_{n-(m-1)\tau,2}, ..., x_{n-(m-1)\tau,N}\right),  \left( x_{n-(m-2)\tau,1}, x_{n-(m-2)\tau,2}, ..., x_{n-(m-2)\tau,N}\right), ... \left(x_{n,1}, x_{n,2}, ..., x_{n,N}\right)$}
\end{pmatrix}
\end{equation}
\subsubsection{Time Delay($\tau$)}\label{section:time-delay}
For practical purpose it is important to compute the appropriate value of the the delay($\tau$) in the first place. For this we had a multidimensional time series in which we computed a multidimensional mutual information and used it's first minima(and global minima, in case the first minima doesn't exist) in a plot between time delay and mutual information.

Let the time series be $x_{n}$ having length N
The time delayed versions are given by
\begin{equation}
x_{n}^{(0)}= x_{n}[1: N-\tau]
\end{equation}
\begin{equation}
x_{n}^{(\tau)}= x_{n}[\tau: N]
\end{equation}
Let the probability distribution functions be $P(x_{n}^{(0)})$, $P(x_{n}^{(\tau)})$ and the joint probability function be $P(x_{n}^{(0)},x_{n}^{(\tau)})$
The entropy measures for each of the signals are given by:
\begin{equation}
H(x_{n}^{(0)})= \sum_{i=1}^{N-\tau} P(x_{n,i}^{(0)}) log(P(x_{n,i}^{(0)}))
\end{equation}

\begin{equation}
H(x_{n}^{(\tau)})= \sum_{i=1}^{N-\tau} P(x_{n,i}^{(\tau)}) log(P(x_{n,i}^{(\tau)}))
\end{equation}
Now let's consider conditional entropy between the two time series. For simplicity now on we would assign $X=x_{n}^{(0)}$ and $Y=x_{n}^{(\tau)}$, and x, y be the individual points which is in the probability distribution functions
\begin{equation}
H(Y|X)= \sum_{x \in X} \sum_{y \in Y} P(x,y)log(\frac{P(x,y)}{P(x)}) = 
\sum_{x \in X} \sum_{y \in Y} P(x,y)[log(P(x,y)-log(P(x))] 
\end{equation}
This gives: 
\begin{equation}
H(Y|X)=\sum_{x \in X} \sum_{y \in Y} P(x,y)log(P(x,y)-\sum_{x \in X} \sum_{y \in Y} P(x,y)log(P(x))
\end{equation}
Second part of LHS would become
\begin{equation}
\sum_{x \in X} \sum_{y \in Y} P(x,y)log({P(x)})= \sum_{x \in X} P(x)log({P(x)})= H(X)
\end{equation}
and first part is:
\begin{equation}
\sum_{x \in X} \sum_{y \in Y} P(x,y)log(P(x,y) = H(X,Y)
\end{equation}
Which gives us
\begin{equation}
H(Y|X)=H(X,Y)-H(X)
\end{equation}
Now let's consider equation for mutual information:
\begin{equation}
I(X,Y)= H(Y) - H(Y|X)
\end{equation}
Using (14) we have
\begin{equation}
I(X,Y)= H(Y) - [H(X,Y) - H(X)] = H(X)+H(Y) - H(X,Y)
\end{equation}
Which we used to compute mutual information. But, we were using multidimensional histogram function(histogramdd function in numpy library) which may encounter overflow error as the number of bins itself can go beyond the maximum number that can get represented in a 64bit binary system. Most Windows systems would encounter this issue, hence it could be difficult as a method(one would simply get a large negative value instead of a positive value). One design choice would be to reduce the number of participants in the group(which attributes to dimensions). We ran the code in a Linux server where it is possible to use 128 bits format to overcome the ‘overflow’ error that one may encounter in typical system configuration on a Windows OS. We used ASUS ESC4000 G3 machine running on Ubuntu 20.04 with Intel Xeon Processor E5-2620 v4(32(8 core), 20M cache, 2.10GHz, turbo upto 3.0GHz), with 128GB DDR4 RAM).

\subsubsection{Embedding Dimension(m)}
For computing the embedding dimension we computed the concept of false nearest neighbours(FNN)(\citeauthor{kennel1992determining}, \citeyear{kennel1992determining}). This method is used to determine the minimum number of false embedding dimensions required to properly reconstruct the attractor. The idea being a properly unfolded attractor would have a large number of false nearest neighbours. These points are those which appear together due to trajectory crossing when we project to smaller dimension. Thus as we increases the embedding dimension those points won't remain as neighbours, suggesting that the embedding dimension we had chosen is not sufficiently unfolding the attractor. In other words the points would be neighbours in the current embedding dimension but, when it is increased they won't as their future temporal dynamics would be too different.

In simple terms, suppose the correct embedding dimension for some time series is $m_{0}$, then when we are eliminating one dimension, these points will get strongly affected by such elimination will become FNN. For this we took the closest neighbours of points in the m dimension and computed the distance and done the same in m+1 dimension and then took the ratio, as given below:
\begin{equation}
X_{fnn}(r)= \frac{\sum_{n=1}^{N-m-1} \Theta(\frac{|x_n^{(m+1)}-x_{k(n)}^{(m+1)}|}{|x_n^{(m)}-x_{k(n)}^{(m)}|}-r) \Theta(\frac{\sigma}{r}-|x_n^{(m)}-x_{k(n)}^{(m)}|)}{\sum_{n=1}^{N-m-1}\Theta(\frac{\sigma}{r}-|x_n^{(m)}-x_{k(n)}^{(m)}|)}
\end{equation}

Where $x_{k(n)}^{(m)}$ is the closest neighbour of $x_n^{(m)}$ in m dimensions, and $\Theta(x)$ is the Heaviside step function. $\sigma$ is the standard deviation of the data. The function k(n) is given as:
\begin{equation}
k(n)= \{n' \mid |x_n^{(m)}-x_{n'}^{(m)}|\leq |x_n^{(m)}-x_{n''}^{(m)}|, \forall n''\in I(x)-n\}
\end{equation}
where I(x) is the set of all indices of x. $\Theta$ denotes step function. A more appropriate method would be to see the FNN ratio as a function of r (see \citeauthor{kantz2004nonlinear}(\citeyear{kantz2004nonlinear}), section 3.3.1, page 37, figure 3.3))

here we are seeing the FNN ratio as a function of r from m=1(top curve) to m=5(bottom). From this one can see that there would be a saturating curve specially where the FNN is hitting zero. For this we defined the radius at which FNN hist zero as 
\begin{equation}
r_{0}(m)= \{r \mid X_{fnn}(r)<\delta, \forall r\in [r_{min},r_{max}]\}
\end{equation}
Where $[r_{min},r_{max}]$ is the interval where we are searching, which is for m embedding dimensions, and $\delta$ is a small enough number. When we do this for different embedding dimensions we will get a graph like the following: 
\begin{suppfigure} 
    \centering
    \includegraphics[scale=0.4]{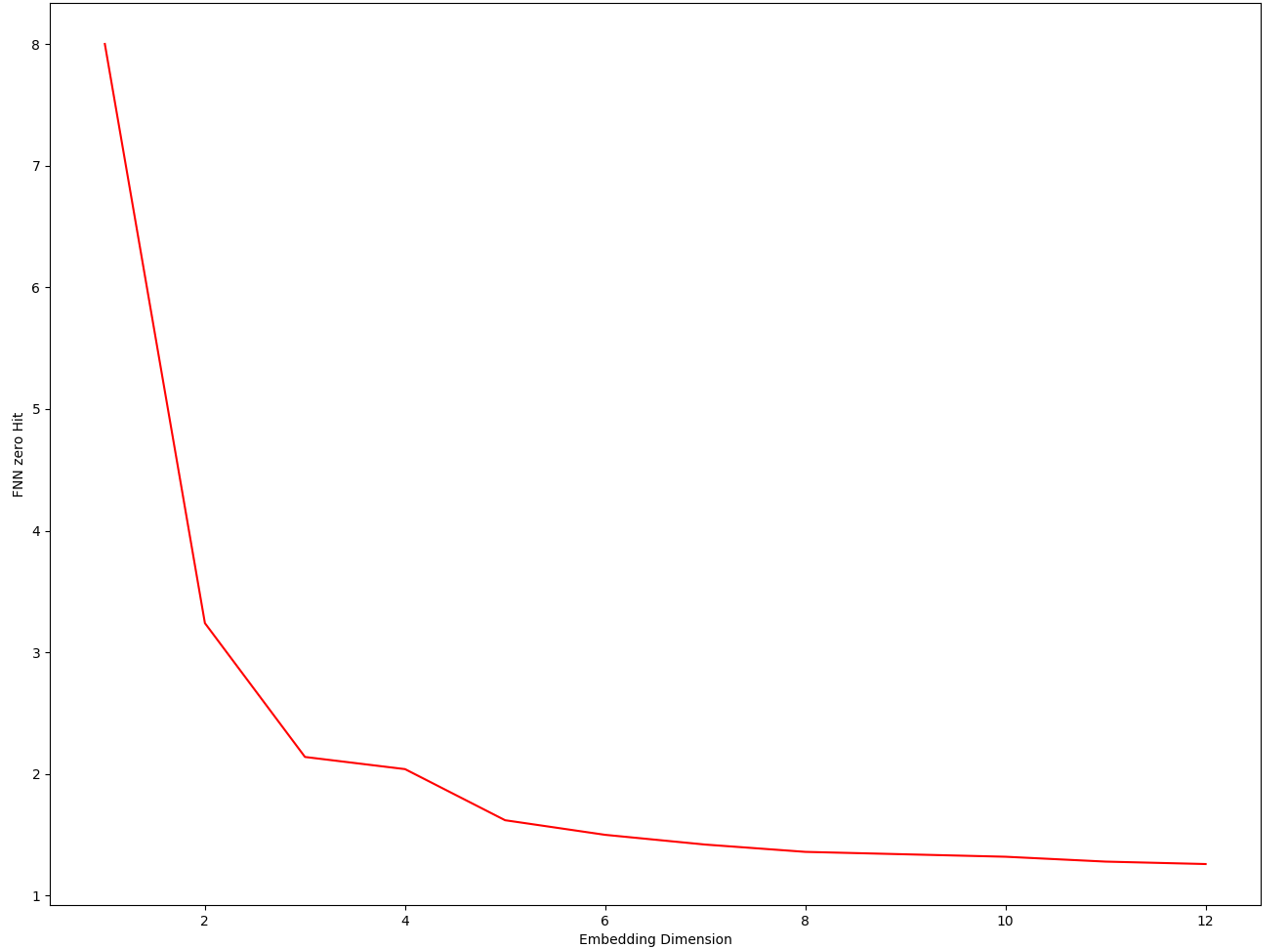}
    \caption{\justifying \footnotesize \textit{$r$ at which FNN hits zero plotted against embedding dimension, we can see that, as we increases the embedding dimension, the FNN hits zero for smaller values of $r$}}
    \label{fig:S2}
\end{suppfigure}

based on this we defined another function, which is as follows: 
\begin{equation}
m=\{max(m') \mid r_{0}(m'-1)-r_{0}(m') \geq \beta \}
\end{equation}

Where if you notice the graph, $\beta$ is a tolerance criteria to stop. In simple words, we will start from the maximum embedding dimension and compare it to the dimension which lack one axis in terms of radius at which FNN is hitting zero and do it until the difference is greater than a bound denoted by $\beta$(=0.2 is what we have chosen). In other words this is a form of exploration to find appropriate embedding dimension. 

\par For finding embedding dimension (m): Since the default tolerance value of 0.2 did not work for arriving at a parameter value in case of many dyads, we explored multiple values of bounds and picked the one that allowed us to estimate m for 97 percentage of the data (excluding data from 9 out of 261 dyads).

\section{Selecting Appropriate Window Size}
\renewcommand{\theequation}{SC.\arabic{equation}}

Recurrent plots are all those points in the trajectory such that they are within a neighbourhood, defined by radius $\epsilon$. In simple terms, let $S_{n}$ be the set of all points in the phase space, then the recurrent plot is given by
\begin{equation}
RP(i,j)= \Theta(\epsilon - || \bar{X}_{i} - \bar{X}_{j}||) 
\end{equation}

Here it is important to note that the selection of appropriate m, $\tau$ and $\epsilon$ is important here. We choose epsilon by setting the recurrence rate to 10 percent independently on each RP.

Since we have differently sized RPs we cannot compare them directly. For this we used sliding windows with step size=1 on each RP and computed representative statistics(mean, mode and median) from these. But, the window size we have chosen should be large enough for capturing the non-linearity. For this we followed a bootstrapping method suggested in \citeauthor{marwan2013recurrence}(\citeyear{marwan2013recurrence}). Let's say we have N number of windows obtained using the sliding window approach. Each of these windows will have histogram distribution of line lengths, and each window represent dynamics at a time point. Let these histogram distributions at t as a function of line length be given by $P_{t}(l)$. Then a unified distribution can be obtained as follows:
\begin{equation}
P(l)= \sum_{t} P_{t}(l)
\end{equation}
The average number of drawings from these histogram distribution will be given as 
\begin{equation}
\overline{n}= \frac{1}{N_{windows}} \sum_{t}\sum_{l} P_{t}(l)
\end{equation}
Where the function $P_{t}(l)$ is actually giving the counts, and $N_{windows}$ is the total number of windows. We will now sample from the histogram distribution for getting $\overline{n}$ number of samples. The probability distribution function is given by:
\begin{equation}
PDF(l)= \frac{P(l)}{\sum_{l} P(l)}
\end{equation}
From this we will compute the cumulative distribution function:
\begin{equation}
CDF(l)= \sum_{i=1}^{l} PDF(i)
\end{equation}
Now we will random sample from this distribution by randomly drawing a number from uniform distribution [0,1] and determining to which line length that number belongs to in terms of CDF.
Let the sample that we are drawing be $I_{i}$
\begin{equation}
I_{i}=r.v(U(0,1))
\end{equation}

where r.v means random variable and U(0,1) is uniform distribution between 0 and 1. 
Then we can apply the inverse of CDF function to get the line length that is corresponding to that particular CDF value
\begin{equation}
I_{i}=CDF(l_{i}) \implies l_{i}=CDF^{-1}(I_{i})
\end{equation}
We repeat this $\overline{n}$ times to get one sample in the bootstrapping. 
\begin{equation}
l=\{ l_{1}, l_{2}, l_{3}, ... , l_{\overline{n}} \}
\end{equation}
We will compute quantitative statistics such as mean lengths, determinism and entropies from this set. and that constitutes one bootstrapping sample. 

We will repeat this sampling 1000 times to construct a distribution of nonlinear measurements for each different window size. From that we computed the difference between 95 percentile quantile and 5 percent quantile as a measure of variance, and found that the decrease is pretty low so between window size 60 and 70, and based on a tolerance criteria we pick window size 68. Then we computed RQA variables using a sliding window having size 68. Provided below is the graph showing the decrease for percentage determinism.

\begin{suppfigure} 
    \centering
    \includegraphics[scale=0.4]{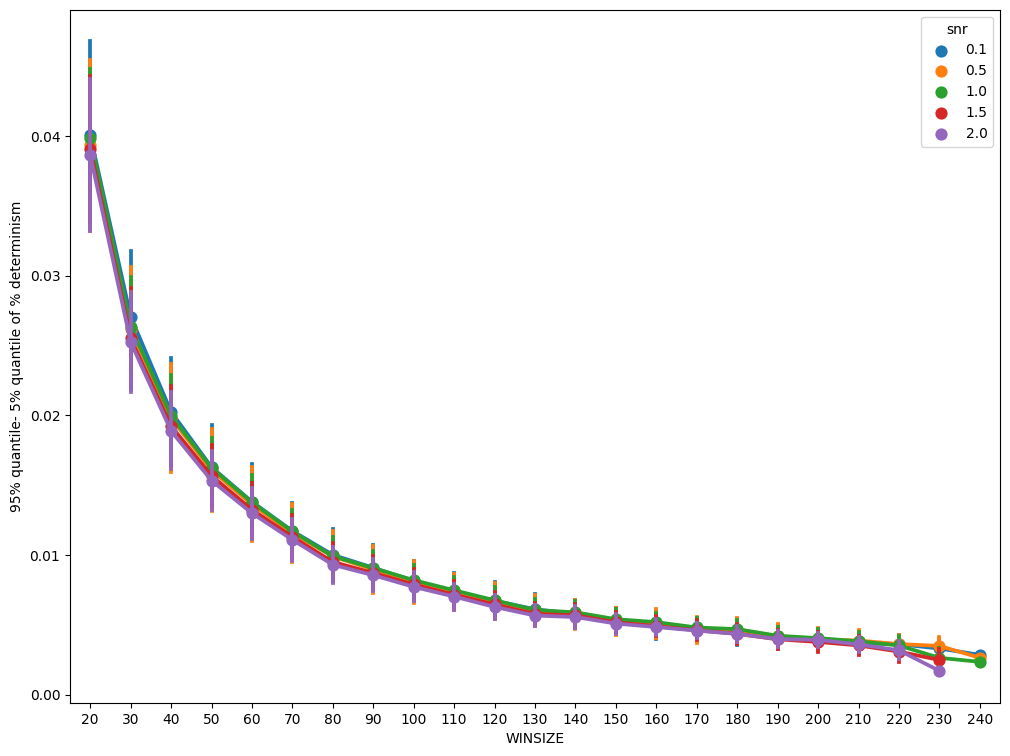}
    \caption{\justifying \footnotesize \textit{The difference between 95th quantile and 5th quantile, plotted against the window size.}}
    \label{fig:S3}
\end{suppfigure}

\section{Nested Cross Validation}
\renewcommand{\theequation}{SD.\arabic{equation}}
\begin{algorithm}[H]
\SetAlgoLined
\SetKwInOut{Input}{Input}
\SetKwInOut{Output}{Output}

\Input{Dataset, Model information, total set of features \& iterations}
\Output{Set of best features, list of performance scores on validation set}

    \For{each fold in the outer loop}{
        Split data into training and validation sets\;
        Initialize best performance score and best feature subset\;
        \For{each feature subset(combination)}{
            \For{each fold in inner loop}{
            Split the training data further into training(train) and validation(train) sets\;
            Train model on the training(train) set with feature subset\;
            Calculate the performance score on the validation set for the current inner fold\;
            Store the performance score for the current inner fold\;
            }
            Calculate aggregate performance score for the feature subset(combination)\;
            \If{aggregate performance score is better than best performance score}{
                Update best performance score\;
                Update best feature subset\;
            }
        }
        Store the best feature for the current outer fold\;
        Train the model on training data with best feature subset for the current outer fold\;
        Calculate the performance score on the validation set for the current outer fold\;
        Store the performance score for the current outer fold\;
    }

\caption{Nested Cross-validation with Best Subset Selection}
\end{algorithm}

\justifying \par The nested cross validation is used mainly to overcome the issue of data leakage. In cross validation, we can't use selected features, as we are dividing the data multiple times into training and validation sets. To avoid data leakage, the feature selection should happen exclusively from the training set and this should be true for each of the training set division. For this, in nested cross validation, an inner loop is used within a loop that loops over all combination of features, to select features from the training set, which would be used to train model only on that training fold, which will be used to estimate the cross validation performance on the validation set for the corresponding iteration. 

\subsection{Comparison of Datasets}
After nested cross validation, what we would have is a list of scores, and we expect using a non parametric test, such as the Wilcoxon test and Sign/Binomial test to be the best method to compare between them. 
\subsubsection{Wilcoxon Sign Rank Test}
Let, the datasets we have be A and B. Suppose, we have N iterationd of a repeated cross validation and $d_{i}$ be the difference between the dataset A compared to B on $i^th$ iteration, given that the same statistical model with same paramaters are used. Differences are ranked based on their values and average ranks are given in case of ties. Let, $R^{A}$ be the sum of ranks, for cases in which performance of the model on dataset A is better than that of B. 
\begin{equation}
    R^{A} = \sum_{d_{i}>0} rank(d_{i}) + \frac{1}{2} \sum_{d_{i}=0} rank(d_{i})
\end{equation}
The $R^{B}$ be the sum of ranks, for cases in which performance of the model on dataset B is better than that of A.
\begin{equation}
    R^{B} = \sum_{d_{i}<0} rank(d_{i}) + \frac{1}{2} \sum_{d_{i}=0} rank(d_{i})
\end{equation}

Then, let's define $T$ as the minimum of this two sums.
\begin{equation}
    T = \min{(R^{A},R^{B})}
\end{equation}
For a large enough $N$, the z-statistic can be defined as: 
\begin{equation}
    z= \frac{T- \frac{1}{4}N(N+1)}{\sqrt{\frac{1}{24}N(N+1)(2N+1)}}
\end{equation}
\subsection{Sign Test}
Here, we are considering in each iteration, whether one dataset is performing better than the other one or not. The test statistic is simply the fraction of times $d_{i}$ was larger than zero. 
\begin{equation}
    P=\frac{\sum_{d_{i}>0}1}{\sum_{i=1}^{N}1}
\end{equation}
With a large enough $N$ normal approximation can be applied to the binomial distribution, where:
\begin{equation}
    \mu = N/2
\end{equation}
\begin{equation}
    \sigma = \frac{\sqrt{N}}{2}
\end{equation}
Depending on the hypothesis, the test can either be one sided or two sided.

\section{Additional Observations From \citeauthor{koul2023interpersonal}
(\citeyear{koul2023interpersonal})}
\newpage
\begin{suppfigure} 
    \centering
    \includegraphics[scale=1.0]{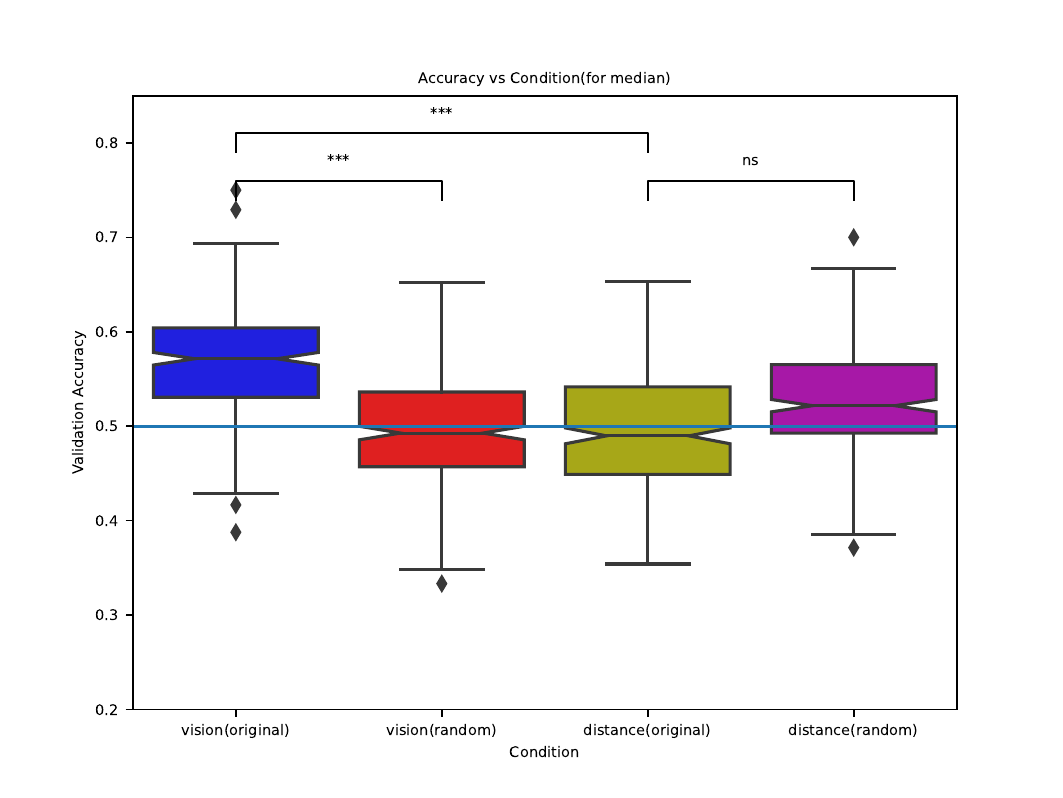}
    \caption{\justifying \footnotesize \textit{Classifier performance measures (Accuracy) using median of the MdRQA variable distributions across sliding windows for predicting visual access ON vs. OFF using the original vs. randomized time series and for predicting Near vs. Far using the original vs. randomized time series.}}
    \label{fig:S4}
\end{suppfigure}
\newpage
\begin{suppfigure} 
    \centering
    \includegraphics[scale=1.0]{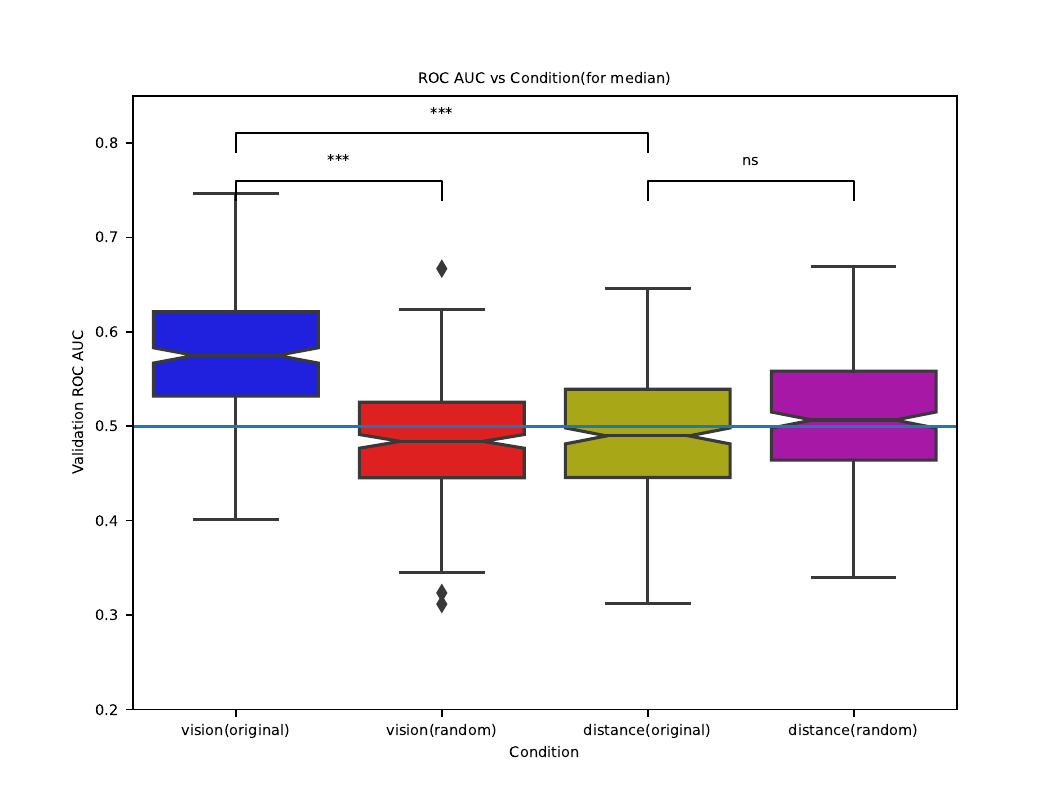}
    \caption{\justifying \footnotesize \textit{Classifier performance measures (ROC AUC) using median of the MdRQA variable distributions across sliding windows for predicting visual access ON vs. OFF using the original vs. randomized time series and for predicting Near vs. Far using the original vs. randomized time series.}}
    \label{fig:S5}
\end{suppfigure}
\newpage
\begin{suppfigure} 
    \centering
    \includegraphics[scale=1.0]{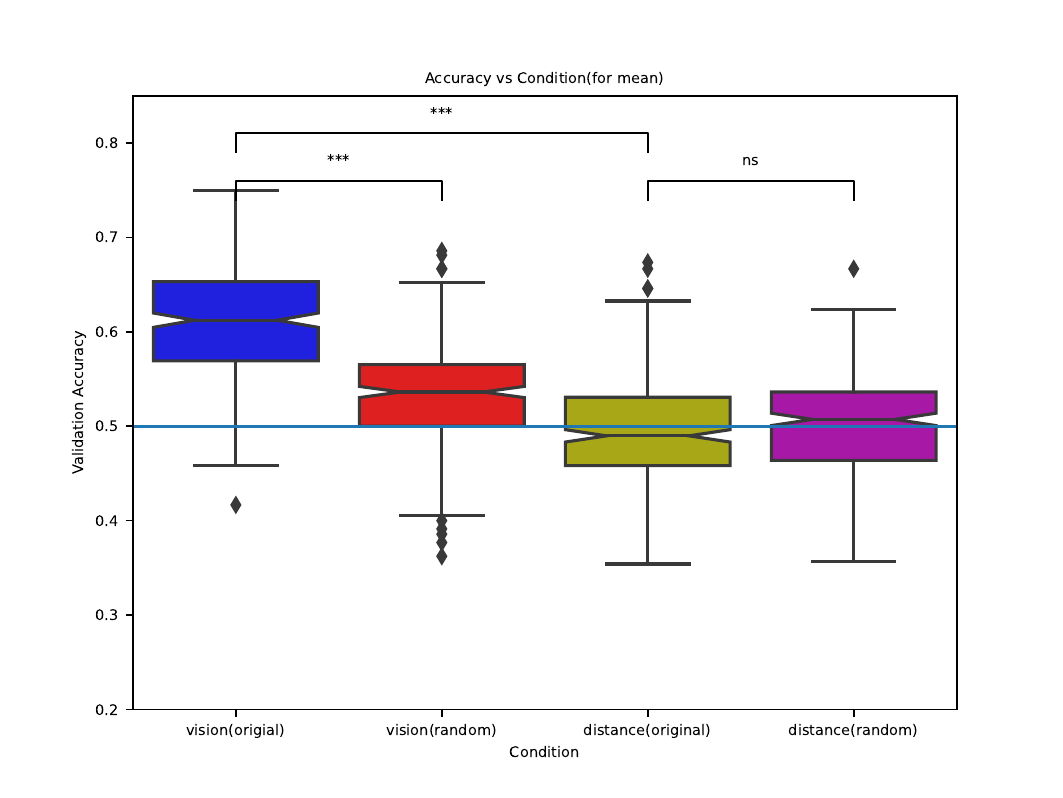}
    \caption{\justifying \footnotesize \textit{Classifier performance measures (Accuracy) using mean of the MdRQA variable distributions across sliding windows for predicting visual access ON vs. OFF using the original vs. randomized time series and for predicting Near vs. Far using the original vs. randomized time series.}}
    \label{fig:S6}
\end{suppfigure}
\newpage
\begin{suppfigure} 
    \centering
    \includegraphics[scale=1.0]{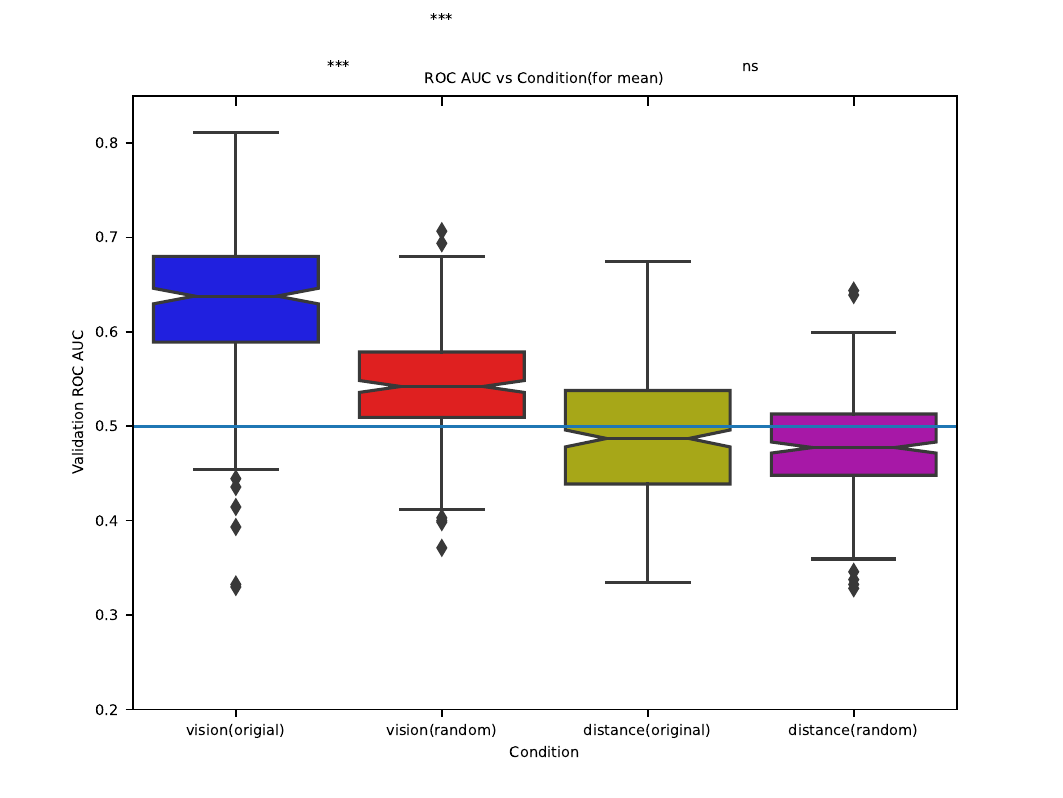}
    \caption{\justifying \footnotesize \textit{Classifier performance measures (ROC AUC) using mean of the MdRQA variable distributions across sliding windows for predicting visual access ON vs. OFF using the original vs. randomized time series and for predicting Near vs. Far using the original vs. randomized time series.}}
    \label{fig:S7}
\end{suppfigure}
\newpage
\begin{suppfigure} 
    \centering
    \includegraphics[scale=1.0]{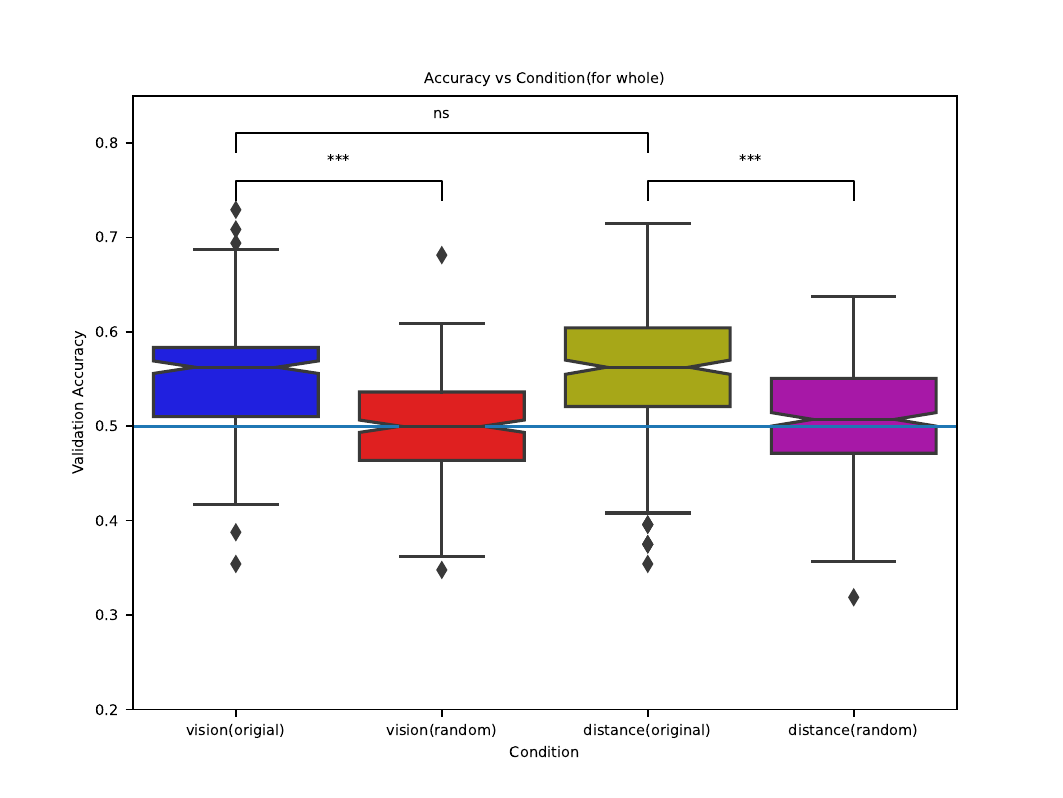}
    \caption{\justifying \footnotesize \textit{Classifier performance measures (Accuracy) using the MdRQA variables of the single RP of the whole time series for predicting visual access ON vs. OFF using the original vs. randomized time series and for predicting Near vs. Far using the original vs. randomized time series.}}
    \label{fig:S8}
\end{suppfigure}
\newpage
\begin{suppfigure} 
    \centering
    \includegraphics[scale=1.0]{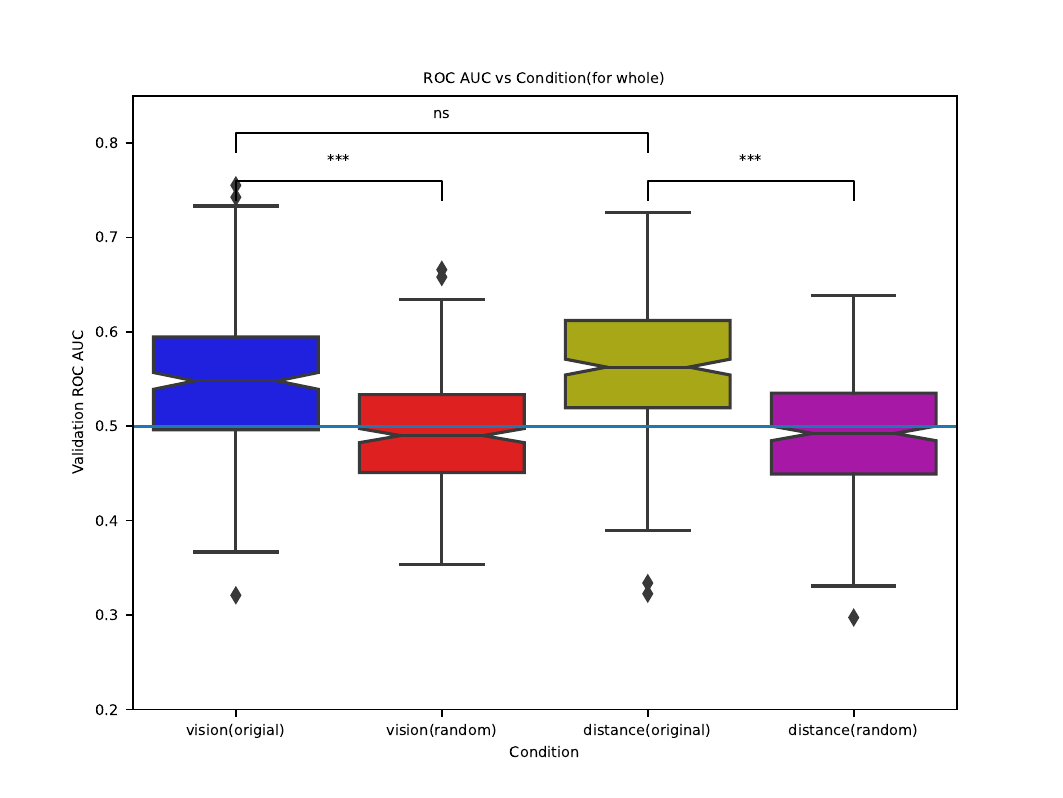}
    \caption{\justifying \footnotesize \textit{Classifier performance measures (ROC AUC) using the MdRQA variables of the single RP of the whole time series for predicting visual access ON vs. OFF using the original vs. randomized time series and for predicting Near vs. Far using the original vs. randomized time series.}}
    \label{fig:S9}
\end{suppfigure}
\newpage
\begin{suppfigure} 
    \centering
    \includegraphics[scale=1.0]{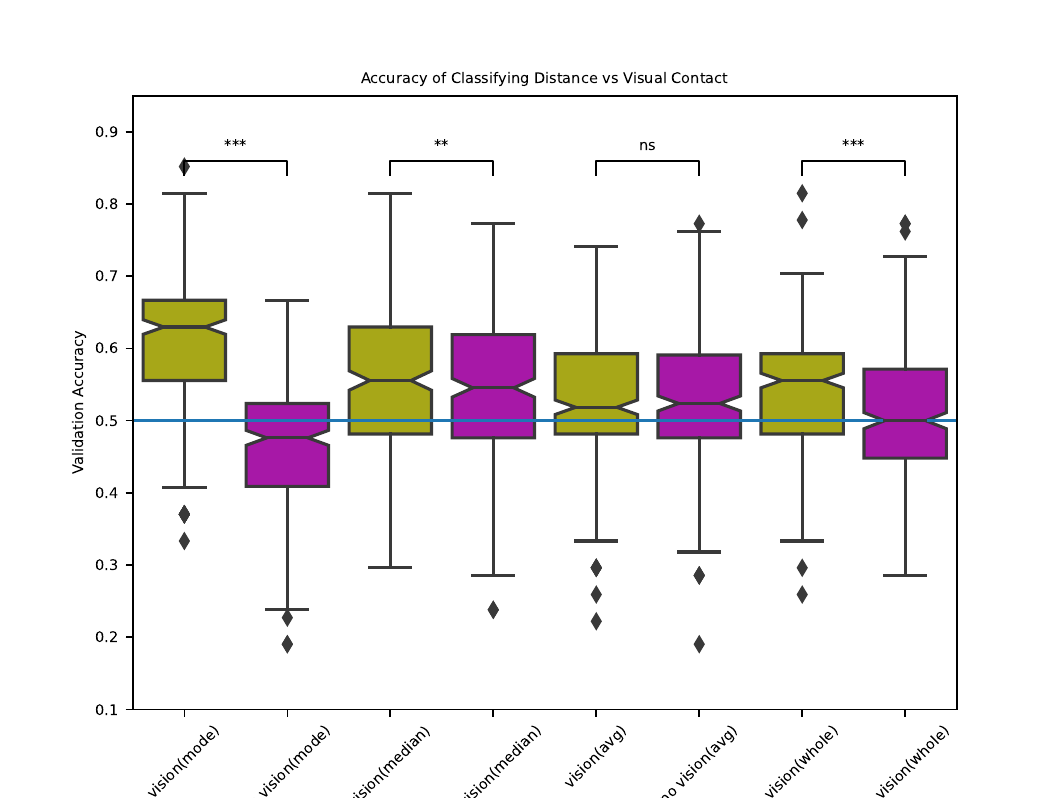}
    \caption{\justifying \footnotesize \textit{Classifier performance measures (Accuracy) for predicting Near vs. Far using trials with visual access ON (yellow) vs. OFF(magenta) using different measures of central tendency (mode, mean, median) of the MdRQA variable distribution across sliding windows as well as the MdRQA variables from the single RP of the whole time series.}}
    \label{fig:S10}
\end{suppfigure}
\newpage
\begin{suppfigure} 
    \centering
    \includegraphics[scale=1.0]{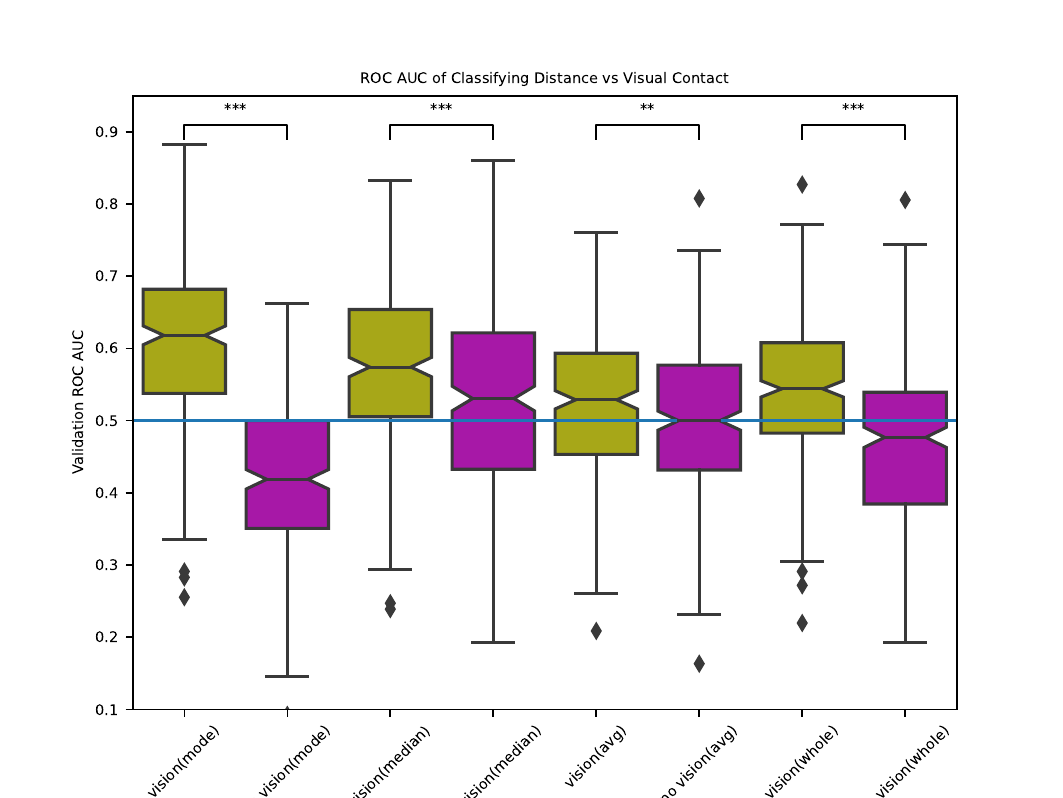}
    \caption{\justifying \footnotesize \textit{Classifier performance measures (ROC AUC) for predicting Near vs. Far using trials with visual access ON (yellow) vs. OFF(magenta) using different measures of central tendency (mode, mean, median) of the MdRQA variable distribution across sliding windows as well as the MdRQA variables from the single RP of the whole time series.}}
    \label{fig:S11}
\end{suppfigure}

\section{ Analysis Pipeline for Sliding Window MdRQA}
\subsection{Test for Nonlinearity}
\par First step is to check some properties of the data for making sure that the data is suitable for a non-linear analysis. It is important to justify the motivation to use a nonlinear method over a linear one, given the computational complexity and computational cost. At the same time, using a linear method without testing for nonlinearity may also lead us to miss some insights. For this we would use \textbf{surrogate analysis} to check whether there is sufficient evidence for opting for a nonlinear analysis. 
\subsubsection{Surrogate Analysis}
\par In surrogate analysis, the null hypothesis is that the data is drawn from a stationary Gaussian process. 
\begin{itemize}
    \item \textbf{Selecting Appropriate Surrogate:} It is important to choose the surrogate carefully to avoid rejecting the null hypothesis by mistake. \citeauthor{kugiumtzis1999test}(\citeyear{kugiumtzis1999test},  \citeauthor{schreiber1996improved}(\citeyear{schreiber1996improved}). 
    \item \textbf{Choosing appropriate measures of nonlinearity:} Choice of a nonlinearity measure depends on what we test and what kind of data we have. In some studies, we may have relatively short stretch of data, where it will be difficult to compute estimates from delayed embedding (e.g. correlation dimension where we may not get a linear region in correlation sum plot). In that case, time irreversibility (\citeauthor{schreiber1997discrimination}, \citeyear{schreiber1997discrimination}) could be suggested as it is computationally cheap and doesn't require delayed embedding. This measure isn't a necessary but is a sufficient condition for non-linearity. \citeauthor{schreiber1997discrimination}(\citeyear{schreiber1997discrimination}). 
\end{itemize}

\subsection{Choose Appropriate \textbf{$r$} value}
In our analysis, we selected an embedding dimension after doing a parameter exploration, after estimating the \textbf{$r$} value at which FNN would hit zero. It is generally suggested to use $\textbf{r} = 10 $. However, it is possible for some dataset that, the FNN values for even small embedding dimensions may hit zero before r = 10, making the embedding dimension computed using this value less useful. A general suggestion is to do a parameter exploration. 

\renewcommand\appendixname{Appendix}

\appendix*

\justifying 
\newpage
\thispagestyle{empty} 

\title{Appendix}
\section{Proof For Volume Shrinking}
\renewcommand{\theequation}{AA.\arabic{equation}}

At this point the reader may have a doubt about why the drop in FNN is faster in the higher dimension. For that we need to understand that the behaviour of measures and metric changes in the higher dimension. And it is important to notice that we are using Euclidian distance for finding the neighbours. In this was one will be able to imagine a spherical region around a point which could be defined as a neighbourhood. Then it is simple enough to look about what actually happens to a hypersphere as we increase the dimension. For simplicity, let's consider a hypersphere in N dimension whose center is in the origin and is defined as: 
\begin{equation}
\sum_{i=1}^{N} x_{i}^{2} =r^{2}
\end{equation}
We can get the volume of hypersphere at Nth dimension by integrating the volume of the hypersphere at N-1 dimension as follows
\begin{equation}
V_{N}(r) = \int_{-r}^{r} V_{N-1} (\sqrt{r^{2}-x^{2}}) dx
\end{equation}
We will show the effect of increase in dimension on the hypersphere volume by generalizing using a few initial examples

Case 1 : In One Dimension

It's a line, hence $V_{1} = 2r$

Case 2: In two dimensions
\begin{equation}
V_{2}(r) = \int_{-r}^{r} V_{1} (\sqrt{r^{2}-x^{2}}) dx = \int_{-r}^{r} 2r (\sqrt{r^{2}-x^{2}}) dx = 2  \int_{-r}^{r}  2(\sqrt{1-(\frac{x^{2}}{r^{2}})}) dx
\ = 2r^{2} \int_{-r}^{r}  (\sqrt{1-(\frac{x}{r}})^{2}) d(\frac{x}{r})
\end{equation}
Given $ \frac{x}{r} = sin(\theta)$ the equation now becomes
\begin{equation}
V_{2}(r)= 2 r^{2} \int_{\frac{-\pi}{2}}^{\frac{\pi}{2}} \sqrt{ 1 - sin(\theta)^{2}} d(sin(\theta))
= 2 r^{2} \int_{\frac{-\pi}{2}}^{\frac{\pi}{2}} cos(\theta)^{2} d(\theta) = \pi r^{2}
\end{equation}

and for higher dimensions
N=3, $V_{3} = \frac{4}{3} \pi r^{3}$
N=4, $V_{4} = \frac{\pi^{2}}{2} r^{4}$

now we have seen
\begin{equation}
V_{N}= k_{N} r^{N}
\end{equation}
Let's proceed to higher dimensions Here note that $k_{N}$ is a variable independent of r.

Now let's consider the volume of hypershpere in Nth dimension in terms of that in N-1th dimension. 

\begin{equation}
V_{N}(r) = \int_{-r}^{r} V_{N-1} (\sqrt{r^{2}-x^{2}}) dx = K_{N-1} \int_{-r}^{r} (r^{2}-x^{2})^{\frac{N-1}{2}} dx = K_{N-1} r^{N} \int_{-r}^{r} (1-(\frac{x}{r})^{2})^{\frac{N-1}{2}}d(\frac{x}{r})
\end{equation}
From this, using trignometry, taking $\frac{x}{r}=sin(\theta)$ we would get: 
\begin{equation}
V_{N}(r) = K_{N-1} r^{N} \int_{-\frac{\pi}{2}}^{\frac{\pi}{2}} cos(\theta)^{N-1} cos(\theta) d(\theta) = K_{N-1} r^{N} \int_{-\frac{\pi}{2}}^{\frac{\pi}{2}} cos(\theta)^{N} d(\theta) 
\end{equation}

From this we can get
\begin{equation}
K_{N} = K_{N-1} \int_{-\frac{\pi}{2}}^{\frac{\pi}{2}} cos(\theta)^{N} d(\theta) = K_{N-1} C_{N}
\end{equation}

This actually leads to a recursive equation if you elaborate $K_{N-1}$
\begin{equation}
K_{N}= C_{N} K_{N-1}= C_{N} C_{N-1} K_{N-2}= C_{N} C_{N-1} C_{N-2} K_{N-3}
\end{equation}
This countinues until N=1, in short we can write the equation as follows
\begin{equation}
K_{N} = K_{1} \prod_{i=2}^{N} C_{i}
\end{equation}

We have 
\begin{equation}
C_{N}= \int_{-\frac{\pi}{2}}^{\frac{\pi}{2}} cos(\theta)^{N} d(\theta) = \int_{-\frac{\pi}{2}}^{\frac{\pi}{2}} cos(\theta)^{2} cos(\theta)^{N-2} d(\theta) =\frac{N-1}{N} \int_{-\frac{\pi}{2}}^{\frac{\pi}{2}}  cos(\theta)^{N-2} d(\theta)= \frac{N-1}{N} C_{N-2}
\end{equation}

similarly we would get 
\begin{equation}
C_{N-1} = \frac{N-2}{N-1} C_{N-3}
\end{equation}
From this
\begin{equation}
C_{N} C_{N-1} = \frac{N-1}{N} C_{N-2} \frac{N-2}{N-1} C_{N-3} = \frac{N-2}{N} C_{N-2} C_{N-3}
\end{equation}

Again this is recursive, if we continue this, we would get: 
\begin{equation}
C_{N} C_{N-1} = \frac{N-2}{N} C_{N-2} C_{N-3}= \frac{N-2}{N}  \frac{N-4}{N} C_{N-4} C_{N-5}
\end{equation}
Now, if we continue this we would get equations as follows:
\begin{equation}
C_{N} C_{N-1}=
\begin{cases}
     \text{ $\frac{2}{N} C_{2}C_{1}$ if $N$ is even}\\
     \text{$\frac{1}{N} C_{0}C_{1}$ if $N$ is odd}\\
\end{cases}
\end{equation}

We know the values of $C_{0}$, $C_{1}$, and $C_(2)$, substitute that in the equation
\begin{equation}
C_{N} C_{N-1}=
\begin{cases}
     \text{ $\frac{2 \pi}{N}$ if $N$ is even}\\
     \text{ $\frac{2 \pi}{N}$ if $N$ is odd}\\
\end{cases}
=\frac{2 \pi}{N}
\end{equation}
We already have 
\begin{equation}
K_{N} = K_{1} \prod_{i=2}^{N} C_{i}
\end{equation}
and 
\begin{equation}
K_{N}= 2 \prod_{i=2}^{N} C_{i}= 
\begin{cases}
     \text{ $2 \frac{2 \pi}{N} \frac{2 \pi}{N-2} ... \frac{2 \pi}{4} C_{2}$ if $N$ is even}\\
     \text{ $2 \frac{2 \pi}{N} \frac{2 \pi}{N-2} ... \frac{2 \pi}{3} $ if $N$ is odd}\\
\end{cases} \\
= 
\begin{cases}
     \text{ $2 \frac{2 \pi}{N} \frac{2 \pi}{N-2} ... \frac{2 \pi}{4} \pi$ if $N$ is even}\\
     \text{ $2 \frac{2 \pi}{N} \frac{2 \pi}{N-2} ... \frac{2 \pi}{3} $ if $N$ is odd}\\
\end{cases}
\end{equation}

From this equation it is easy to see that for a sphere having radius r, as the number of dimension increases the volume decreases. If we take limit of N tending to infinity the term $K_{N}$ and effectively the volume would become zero.

This is the reason why as we increase the radius in higher dimension the false neighbour ratio falls quickly. If we look carefully at the equation for FNN we can notice that it is checking whether the ratio between the distance between two points in Nth and N-1th dimension is higher than a factor(r). But, when you think carefully about the volume shrinking, we can see that at higher dimension, everyone would become neighbour of another one, as the volume of any sphere will shrink towards zero. Hence, the drop would be steeper. Simply it means that your neighbourhood is getting sparse so that all points from you appear as if they are at the same distance from you and you won't be able to sample the entire volume. This is also the reason why selecting an appropriate embedding dimension is important, as a very high dimension will effectively make all points in the phase space neighbours of each other. 
Implication of this concept is same here also. When you go towards higher dimension, due to the volume shrinking, the hypersphere defining your neighbourhood will also shrink. This results in each other point being at an infinite distance from one point, or more correctly at equal distance. The differences between points in terms of neighbourhood will be hence less observable at higher dimension as any sphere defining neighbourhood will converge into a zero volume sphere. This is more commonly referred to as curse of dimensionality in the field of data science or machine learning. 

\end{document}